\documentclass[superscriptaddress,
 amsmath,amssymb,
 aps,
pra,
reprint
]{revtex4-2}

\usepackage{graphicx}
\usepackage{dcolumn}
\usepackage{bm}
\usepackage{physics}
\usepackage{amsfonts, amsmath}
\usepackage{dsfont}
\usepackage{amsthm}
\usepackage{bm}
\usepackage{hyperref}
\usepackage{bbold}
\usepackage{color}
\usepackage{lipsum,babel}
\usepackage[normalem]{ulem}
\usepackage{tabularx}

\newcommand{\cor}[1]{{\color{red}{#1}}}

\newcommand{\haf}{\text{haf}}
\newcommand{\lhaf}{\text{lhaf}}
\newcommand{\per}{\text{Per}}
\newcommand{\poly}{\text{poly}}

\begin{document}
\definecolor{navy}{RGB}{46,72,102}
\definecolor{pink}{RGB}{219,48,122}
\definecolor{grey}{RGB}{184,184,184}
\definecolor{yellow}{RGB}{255,192,0}
\definecolor{grey1}{RGB}{217,217,217}
\definecolor{grey2}{RGB}{166,166,166}
\definecolor{grey3}{RGB}{89,89,89}
\definecolor{red}{RGB}{255,0,0}

\preprint{APS/123-QED}

\title{Quantum-inspired classical algorithm for molecular vibronic spectra}
\author{Changhun Oh}
\affiliation{Pritzker School of Molecular Engineering, University of Chicago, Chicago, Illinois 60637, USA}
\author{Youngrong Lim}%
\affiliation{School of Computational Sciences, Korea Institute for Advanced Study, Seoul 02455, Korea}
\author{Yat Wong}
\affiliation{Pritzker School of Molecular Engineering, University of Chicago, Chicago, Illinois 60637, USA}
\author{Bill Fefferman}
\affiliation{Department of Computer Science, University of Chicago, Chicago, Illinois 60637, USA}
\author{Liang Jiang}
\affiliation{Pritzker School of Molecular Engineering, University of Chicago, Chicago, Illinois 60637, USA}

\begin{abstract}
We have recently seen the first plausible claims for quantum advantage using sampling problems such as random circuit sampling using superconducting qubits and Gaussian boson sampling.
The obvious next step is to channel the potential quantum advantage to solving practical applications rather than proof-of-principle experiments.
Recently, a quantum simulator, specifically a Gaussian boson sampler, has been proposed to generate molecular vibronic spectra efficiently, which is an essential property of molecules and an important tool for analyzing chemical components and studying molecular structures.
Computing molecular vibronic spectra has been a challenging task, and its best-known classical algorithm scales combinatorially in the system size.
Thus, it is a candidate of tasks for which a quantum device provides a computational advantage.
In this work, we propose a quantum-inspired classical algorithm for molecular vibronic spectra for harmonic potential.
We first show that the molecular vibronic spectra problem corresponding to Fock-state boson sampling can be efficiently solved using a classical algorithm as accurately as running a boson sampler.
In particular, we generalize Gurvits's algorithm to approximate Fourier components of the spectra of Fock-state boson sampling and prove using Parseval's relation that the error of the spectra can be suppressed as long as that of the Fourier components are small.
We also show that the molecular vibronic spectra problems of Gaussian boson sampling, which corresponds to the actual molecular vibronic spectra problem in chemistry, can be exactly solved even without Gurvits-type algorithms.
Consequently, we demonstrate that those problems are not candidates of quantum advantage.
We then provide a more general molecular vibronic spectra problem, which is also chemically well-motivated, for which our method does not work anymore and so we might be able to take advantage of a boson sampler.
\end{abstract}

\maketitle

Quantum computers are believed to solve problems that cannot be efficiently solved using classical counterparts \cite{nielsen2002quantum}.~Ultimately, quantum computers need to be fault-tolerant and scalable to solve various problems, such as integer factorization \cite{shor1994algorithms} and digital quantum simulation of the real-time dynamics of large quantum systems \cite{lloyd1996universal}.~While the theoretical study and experimental implementations of quantum error correction schemes are rapidly developing, quantum machines at hand are still noisy intermediate-scale quantum (NISQ) devices.
Nevertheless, we have recently seen the first plausible quantum advantage demonstrations using NISQ devices, such as a superconducting qubit system with random circuit sampling \cite{arute2019quantum, wu2021strong} and a quantum optical system with Gaussian boson sampling \cite{zhong2020quantum, zhong2021phase, madsen2022quantum}.

Since quantum advantages from sampling tasks are promising, there have been proposals for applications for which we can potentially exploit the advantages.~One such example is the generation of so-called molecular vibronic spectra \cite{huh2015boson}.~Due to its importance in chemistry, its classical algorithm has been extensively studied while the best-known algorithm still scales combinatorially in the system size \cite{barone2009vibrationally}.~Recently, it has been proposed to employ a quantum simulator, which is a Gaussian boson sampler \cite{hamilton2017gaussian}, to efficiently generate the spectra \cite{huh2015boson}.~Gaussian boson sampling is a task that is believed to be hard for classical computers under plausible computational complexity assumptions and that has been exploited for a quantum advantage demonstration \cite{aaronson2011computational, hamilton2017gaussian, deshpande2021quantum}.
Therefore, the molecular vibronic spectra problem has been treated as a candidate for applications of quantum simulators \cite{arrazola2021quantum}.~Indeed, many experiments have been conducted to generate spectra using a Gaussian boson sampler \cite{shen2018quantum, paesani2019generation, wang2020efficient}.
Also, many classical algorithms to simulate Gaussian boson sampling have been proposed \cite{quesada2020exact, wu2020speedup, bulmer2021boundary, quesada2022quadratic, oh2022classical}, while the cost is exponential in the system size in general.~Thus, solving the molecular vibronic spectra problem by simulating Gaussian boson sampling takes exponential costs.

Meanwhile, the proposed quantum algorithms might also inspire us to develop efficient classical algorithms \cite{drucker2009quantum}.
For example, the D-wave device has inspired efficient algorithms \cite{shin2014quantum, heim2015quantum}. 
Some quantum machine learning algorithms also inspired novel polynomial scaled classical algorithms \cite{tang2019quantum, gilyen2018quantum}.
More recently, S. Aaronson wrote a blog post claiming that there can be an efficient approximate classical algorithm for molecular vibronic spectra, although he did not give the details \cite{aaronson2020}.

In this work, by using the framework of boson sampling, we establish the complexity of the molecular vibronic spectra problem.~We first consider the problem corresponding to Fock-state boson sampling and show that the exact computation of the spectra is computationally hard (\#P-hard).~However, even running a boson sampler entails a sampling error, which indicates that an additive error, instead of exact computation, is a correct target for classical algorithms.
To achieve the target accuracy as efficient as a boson sampler, we devise an approximate classical algorithm using Fourier transformation.
In particular, we generalize Gurvits's algorithm to approximate the Fourier components of general Fock-state molecular vibronic spectra which include multiphoton input states and show by using Parseval's relation that the generalized Gurvits's algorithm and the inverse Fourier transformation achieves the same accuracy as a boson sampler in an efficient way.
We then generalize the method to an actual molecular vibronic spectra problem corresponding to Gaussian boson sampling and show that we can even exactly compute the Fourier components and spectra efficiently.
Consequently, we demonstrate that the molecular vibronic spectra problem corresponding to Fock-state boson sampling or Gaussian boson sampling does not provide a quantum advantage.
We then show by providing a more general type of molecular vibronic spectra problem that there exists a problem, which is also chemically well motivated, that the proposed method does not solve the problem as precisely as running a corresponding boson sampler.
Thus, it suggests that we might be able to take advantage of boson sampling to solve such a classically hard problem.

\begin{figure*}[t]
\includegraphics[width=380px]{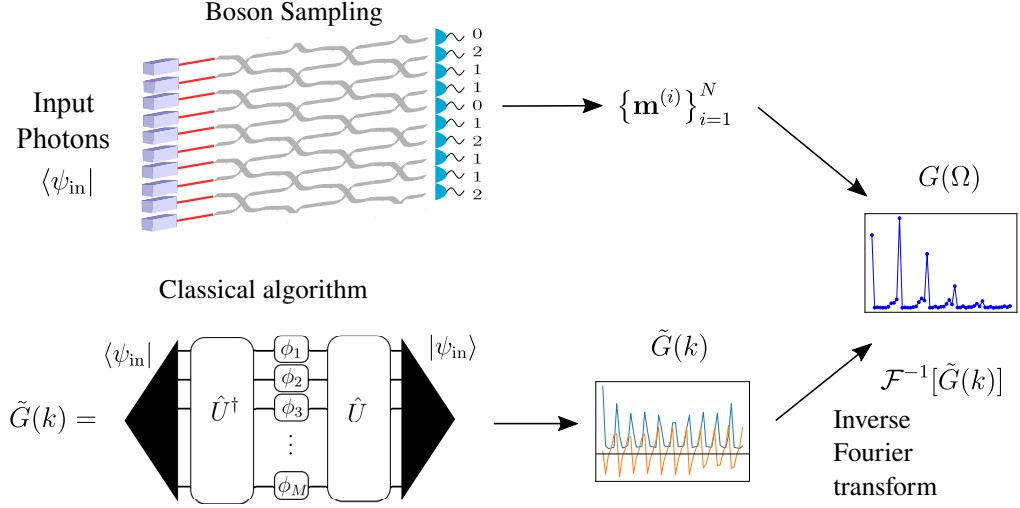} 
\caption{Molecular vibronic spectra generation using boson sampling and the proposed classical algorithm by computing the Fourier components of the spectra.
The Fourier components are described by a circuit diagram whose expression is given in Eq.~\eqref{eq:phase} with defining $\phi_i=-k\theta \omega_i$.
The operation for each $\phi_i$ is phase-shifting operator $e^{i\phi_i\hat{n}_i}$.
}
\label{fig:scheme}
\end{figure*}

\section*{Results}
\subsection*{Grouped probability of boson sampling}
Before we present our main results for the molecular vibronic spectra problem, let us first discuss how to potentially exploit a boson sampling for estimating a computational hard quantity, which provides a more general sense of the molecular vibronic problem.
One naive way of using a boson sampler is to estimate a permanent, which is a hard graph-theoretical quantity, because the output probability of boson sampling can be expressed as a permanent.
A caveat is that, in the hardness regime, the probability, or the corresponding permanent, is exponentially small that we still need exponentially many samples to achieve a reasonable accuracy due to the nature of sampling, even without considering experimental imperfections.
Such a caveat suggests that if we want to estimate a quantity using a boson sampler, it needs to be sufficiently large that the required number of samples is at most polynomially many.
One way to construct such problems is to group output probabilities and estimate their sum, instead of estimating individual probabilities, so the quantity becomes sufficiently large.
The molecular vibronic spectra problem is exactly such kind of a problem, which will be elaborated on in the following section.

\subsection*{Molecular vibronic spectra}
Let us first define the problem using the framework of Fock-state boson sampling and then generalize the problem to the one corresponding to Gaussian boson sampling, which describes the actual molecular vibronic spectra (we define the problem from the chemical perspective and elaborate on the relation in Methods).
Consider an $M$-mode Fock-state boson sampling with input state $|\psi_\text{in}\rangle=|\bm{n}\rangle$ with $\bm{n}\equiv (n_1,\dots,n_M)\in \mathbb{Z}_{\geq 0}^M$ and an $M$-mode linear-optical circuit $\hat{U}$, characterized by $M\times M$ unitary matrix $U$.
The probability of measuring the output photon configuration $|\bm{m}\rangle$ is then written as
\begin{align}
    p(\bm{m})=\frac{|\per U_{\bm{n},\bm{m}}|^2}{\sqrt{\bm{n}!\bm{m}!}},
\end{align}
where $U_{\bm{n},\bm{m}}$ is obtained by repeating the $i$th row of $U$ by $n_i$ times and the $j$th column of $U$ by $m_j$ times.
We then introduce two more parameters which define groups of probabilities:
a weight vector $\bm{\omega}\in \mathbb{Z}^M_{\geq 0}$ with each element at most polynomially large, i.e., $\omega_i\leq O(\text{poly}(M))$ for all $i\in [M]$, and a set of numbers $\Omega\in\{0,\dots,\Omega_\text{max}\}$ which represents each group of outcomes. (The polynomial-size integer weight vector corresponds to the polynomially accurate harmonic angular frequency in spectroscopy. See Methods for more discussion about the weight vector. We lift this assumption below.)
Using the parameters, we group output probabilities in a way that we sum them if the inner product $\bm{\omega}\cdot\bm{m}$ is equal, i.e.,
\begin{align}\label{eq:spectra}
    G(\Omega)
    \equiv \sum_{\bm{m}=\bm{0}}^\infty p(\bm{m})\delta(\Omega-\bm{\omega}\cdot\bm{m})
    =\sum_{\bm{m}\in \mathcal{G}(\Omega)}p(\bm{m}),
\end{align}
where we defined sets $\mathcal{G}(\Omega)\equiv \{\bm{m}\in \mathbb{Z}^M_{\geq 0}:{\bm{m}\cdot\bm{\omega}}=\Omega\}$ for $\Omega\in \{0,\dots,\Omega_\text{max}\}$.
We will call such a grouped probability $G(\Omega)$ as molecular vibronic spectra in a more general sense.
Intuitively, computing the spectra seems difficult because the probability $p(\bm{m})$ is hard to compute and furthermore for each $\Omega$, we need to sum $p(\bm{m})$ over exponentially many outcomes.
Even if we are able to approximate the individual probabilities with a reasonable error, it is nontrivial to approximate the grouped probabilities with suppressing the error because the number of outcomes are exponentially large.

Meanwhile, a boson sampler enables us to straightforwardly estimate the spectra by collecting samples and counting the number of samples that give $\Omega$ after inner product (see Fig.~\ref{fig:scheme}).
Since the number of distinct groups is $\Omega_\text{max}+1\leq O(\text{poly}(M))$, each grouped probability's magnitude $|G(\Omega)|$ becomes $O(1/\text{poly}(M))$; thus, a reasonable target error becomes $O(1/\text{poly}(M))$ as well.
By Chernoff bound, the running time to achieve the target error $\epsilon$ with high probability $1-\delta$ is $T=O(\text{poly}(M,1/\epsilon,\log \delta^{-1}))$, i.e., at most polynomial in the system size and the target accuracy $1/\epsilon$.
We emphasize that in contrast to estimating each probability, which is exponentially small, grouping the probabilities into polynomially many groups allows the sampling error $O(1/\text{poly}(M))$ obtained with polynomially many samples to become still reasonable.

\subsection*{Fourier components of molecular vibronic spectra}
As seen from the definition of the spectra in Eq.~\eqref{eq:spectra}, a direct computation or even an approximation does not seem trivial.
The key idea of our algorithm to circumvent the direct computation of exponentially many probabilities is to consider its Fourier components.
First of all, we derive the Fourier components of molecular vibronic spectra (See Methods for the derivation): 
\begin{align}
    \tilde{G}(k)
    &\equiv \left\langle \bigotimes_{i=1}^M\left(\sum_{m_i=0}^{\infty}|m_i\rangle\langle m_i|_{i}e^{-ik\theta m_i \omega_i}\right)\right\rangle 
    =\left\langle e^{-ik \theta\hat{\bm{n}}\cdot\bm{\omega}}\right\rangle, \label{eq:phase} \\ 
    G(\Omega)
    &=\frac{1}{\Omega_\text{max}+1}\sum_{k=0}^{\Omega_\text{max}}\tilde{G}(k)e^{i k\theta \Omega} 
    =\sum_{\bm{m}=\bm{0}}^{\infty}p(\bm{m})\delta(\Omega-\bm{\omega}\cdot\bm{m}) \label{eq:G},
\end{align}
where $\theta\equiv 2\pi/(\Omega_\text{max}+1)$, $k\in \{0,\dots, \Omega_\text{max}\}$, and $|m_i\rangle\langle m_i|_i$ is the projector on $m_i$ photon state for the $i$th mode.
Note that we did not assume Fock-state boson sampling, i.e., the relation is applicable to any circuits.
Hence, in general, the Fourier component is the overlap between an output state $|\psi_\text{out}\rangle$ and a phase-shifted state $e^{-ik\theta \hat{\bm{n}}\cdot\bm{\omega}}|\psi_\text{out}\rangle$, and once we can compute the overlap efficiently, we obtain the Fourier components and recover the spectra by the inverse Fourier transformation (see Fig.~\ref{fig:scheme}).
Note that the Fourier components depend only on input photons $\bm{n}$ not on output configurations $\bm{m}$.

For Fock-state boson sampling, the Fourier components can be written as
\begin{align}
    \tilde{G}(k)
    =\langle \bm{n}|\hat{U}^\dagger e^{-ik\theta \hat{\bm{n}}\cdot\bm{\omega}}\hat{U}|\bm{n}\rangle
    \equiv \langle \bm{n}|\hat{V}|\bm{n}\rangle
    =\frac{\per(V_{\bm{n},\bm{n}})}{\bm{n}!},
\end{align}
where we defined a unitary operator $\hat{V}\equiv \hat{U}^\dagger e^{-ik \hat{\bm{n}}\cdot\bm{\omega}\theta}\hat{U}$, consisting of $\hat{U}$ and a phase-shift operator $e^{-ik \hat{\bm{n}}\cdot\bm{\omega}\theta}$ and $\hat{U}^\dagger$.
Thus, it is a linear-optical circuit characterized by a unitary matrix $V=U^\dagger D U$, with $D\equiv \text{diag}(e^{-ik\theta \omega_1},\dots,e^{-ik\theta \omega_M})$ characterizing the phase-shift operator.
Here, such a diagonal form suggests that $V$ is a general form of a unitary matrix.
Together with this fact, since computing the permanent of an arbitrary matrix in multiplicative error is \#P-hard and one can embed an arbitrary complex matrix into a submatrix of a unitary matrix by normalizing the matrix \cite{aaronson2011computational}, computing its Fourier component in  multiplicative error is also \#P-hard.
Therefore, it shows that computing Fourier components in multiplicative error is a \#P-hard problem, and consequently, it proves that the exact computation of the spectra is also a \#P-hard problem.



In fact, we again emphasize that even if we run a boson sampling experiment using a quantum device and reproduce the spectra from the sampling outcomes, the resultant spectra has an additive sampling error.
Therefore, reproducing the spectra within a multiplicative error is not expected to be achievable by running a boson sampling circuit with polynomially many samples; thus, an additive error is a relevant target using a classical simulation to compare with a quantum device.
Interestingly, there exists a classical algorithm, so-called Gurvits's algorithm, which can efficiently approximate the permanent of a matrix within an additive error \cite{gurvits2005complexity, aaronson2011computational}.
However, this algorithm and its slightly generalized version \cite{aaronson2012generalizing} are insufficient for general Fock-state boson sampling cases where the input and output photons $n_i$ contain more than a single photon for some $i$'s because the error can increase exponentially in the system size in that case (see Ref.~\cite{supple}).
We further generalize Gurvits's algorithm from Ref.~\cite{gurvits2005complexity} using a similar technique in  Ref.~\cite{aaronson2011computational} to estimate the spectra which includes multiphotons $\bm{n}$ using the following equality (see Ref.~\cite{supple} for the derivation)
\begin{align}
    \frac{\per(V_{\bm{n},\bm{n}})}{\bm{n}!}
    =\mathbb{E}_{\bm{x}\in\mathcal{X}}\left[\prod_{i=1}^M\left(\frac{\bar{y}_i (Vy)_i}{n_i}\right)^{n_i}\right],
\end{align}
where $y_i\equiv \sqrt{n_i}x_i$, $\bm{x}\in \mathcal{X}\equiv \mathcal{R}[n_1+1]\times \cdots \times \mathcal{R}[n_M+1] $, where $\mathcal{R}[j]$ is the set of $j$th roots of unity.
Thus, by sampling the random variable with uniform $\bm{x}\in \mathcal{X}$, the randomized algorithm gives an estimate $\mu$ of permanent of an $n\times n$ matrix such that
\begin{align}
    \left|\frac{\per(V_{\bm{n},\bm{n}})}{\bm{n}!}-\mu\right|<\epsilon\|V\|^n,
\end{align}
with high probability $1-\delta$ with running time $T=O(\text{poly}(n,1/\epsilon,\log\delta^{-1})$).
Here, $\|V\|$ is the spectral norm of the matrix $V$, which is always 1 for our case because $V$ is a unitary matrix.
Thus, the generalized Gurvits's algorithm enables us to efficiently approximate Fourier components within a reasonable additive error even though computation of the Fourier components in a multiplicative error is hard (\#P-hard).
Note that although a generalized Gurvits's algorithm might be used to approximate individual probabilities in Eq.~\eqref{eq:G}, it is nontrivial to approximate the sum of exponentially many probabilities, which is enabled by approximating Fourier components instead of probabilities.

The remaining challenge is the propagation of the error of Fourier coefficients to that of the spectra through the inverse Fourier transformation.
Using Parseval's relation, we prove that as long as we estimate the Fourier coefficients with a small error $\epsilon$, the transformed spectra's error is also small as $\epsilon$ (see Methods for proof):
\begin{align}\label{eq:parseval}
    \sum_{\Omega=0}^{\Omega_\text{max}}|\Delta G(\Omega)|^2=\frac{1}{\Omega_\text{max}+1}\sum_{k=0}^{\Omega_\text{max}}|\Delta \tilde{G}(k)|^2\leq \epsilon^2,
\end{align}
which proves that $|\Delta G(\Omega)|\leq \epsilon$ for any $\Omega$, where $\Delta G$ and $\Delta\tilde{G}$ represent the error of spectra and Fourier component estimation.
Hence, if there is an efficient algorithm that approximates the Fourier components within an error $\epsilon$ in a running time $T=O(\text{poly}(M,1/\epsilon,\log \delta^{-1}))$, the algorithm enables us to achieve the same accuracy as running a boson sampling.
For the Fock-state boson sampling case, the generalized Gurvits's algorithm is such an algorithm estimating Fourier components.
Consequently, we have shown that the molecular vibronic spectra problem corresponding to Fock-state boson sampling can be efficiently solved by a classical computer as accurately as by running a boson sampler, which indicates that there is no quantum advantage from this problem.

Finally, we lift the assumption that the weight vector is at most polynomially large.
In this case, since $\Omega_{\text{max}}=\omega(\poly(M))$, the standard Fourier transformation costs superpolynomial time.
However, notice that even for the boson sampling case, it costs superpolynomial time to estimate all the quantities in a superpolynomial number of bins.
Thus, let us focus on estimating the largest $t=O(\poly(M))$ elements of $G(\Omega)$, namely, peaks, which are large as $\Omega(1/\poly(M))$ to guarantee that a boson sampler can provide a reasonable estimate.
In this case, we show that by using the sparse fast Fourier transformation \cite{hassanieh2012nearly}, we can efficiently approximate the peaks in $O(\poly(M,1/\epsilon))$ with high probability \cite{supple}.
Specifically, we show that such a procedure with approximated Fourier components by the generalized Gurvits's algorithm still enables us to achieve an accuracy as good as running a boson sampler (See Methods and Ref.~\cite{supple}).

\subsection*{Exact computation of Fourier components of actual molecular vibronic spectra}
As mentioned before and in Methods, actual molecular vibronic spectra problems at zero temperature under harmonic potential approximation correspond to Gaussian boson sampling.
For this case where the input state is a product Gaussian state $|\psi_\text{in}\rangle=\hat{D}(\bm{\alpha})\hat{S}(\bm{r})|\bm{0}\rangle$ and the circuit is again a linear-optical circuit $\hat{U}$, the Fourier components can be written as
\begin{align}
    \tilde{G}(k)=\langle \psi_\text{in}|\hat{U}^\dagger e^{-ik\theta\hat{\bm{n}}\cdot \bm{\omega}}\hat{U}|\psi_\text{in}\rangle.
\end{align}
Since only a Gaussian state and Gaussian operations are involved in Fourier components, one can easily compute them using the standard technique of quantum optics without needing Gurvits's approximate algorithm.
More explicitly, we use the positive $P$-representation \cite{drummond1980generalised} for Eq.~\eqref{eq:phase}, motivated by Ref.~\cite{drummond2021simulating}.

\begin{figure}[t]
\includegraphics[width=240px]{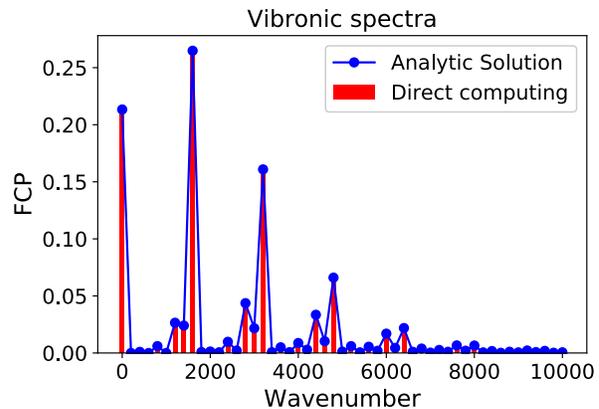}
\caption{Molecular vibronic spectra of formic acid ($\text{CH}_2\text{O}_2, 1^1\text{A}'\to 1^2\text{A}'$) generated by directly computing all probabilities and by the solution in Eq.~\eqref{eq:solution}. The unit of wavenumber is $\text{cm}^{-1}$. The resolution for the spectra is set to be $200~\text{cm}^{-1}$. 
We first compute Fourier components using Eq.~\eqref{eq:solution} and take the inverse Fourier transform to obtain the spectra.
The corresponding transformation is obtained from Ref.~\cite{huh2015boson}.}
\label{fig:ideal}
\end{figure}

Using the positive $P$-representation of single-mode Gaussian states and the relation for a normal-ordered operator such as $e^{-\gamma\hat{n}}=:e^{\hat{n}(e^{-\gamma}-1)}:$ \cite{vargas2006normal}, we can rewrite the Fourier components in Eq.~\eqref{eq:phase} as \cite{supple}
\begin{align}
    \tilde{G}(k)
    &=\int_{\mathbb{R}^{2M}}d\bm{x}d\bm{y}P_\text{in}(\bm{x},\bm{y})\exp\left[\sum_{i=1}^Mx_i'y_i'(e^{-ik\theta \omega_i}-1)\right], \label{eq:p_method}
\end{align}
where $P_\text{in}(\bm{x},\bm{y})$ is the positive $P$-representation of an input squeezed state.
Here, ${:\hat{O}:}$ is the normal-ordered form of an operator $\hat{O}$ \cite{walls2007quantum}, and $(\bm{x}',\bm{y}')=(U\bm{x},U^*\bm{y})$ accounts for the linear-optical unitary operation.
More specifically, when a coherent state goes through an $M$-mode linear-optical circuit $\hat{U}$, it transforms as ${\hat{U}|\bm{x}\rangle=|U\bm{x}\rangle}$, where $U$ is the corresponding $M\times M$ unitary matrix for the circuit.
Since $\tilde{G}(k)$ is now written as a Gaussian integral, it can be analytically obtained, \begin{align}\label{eq:solution}
    \tilde{G}(k)=\mathcal{N}\frac{(2\pi)^M}{\sqrt{\text{det}(Q)}}\exp\left(\frac{1}{2}\bm{c}^\text{T}Q^{-1}\bm{c}+c_0\right),
\end{align}
where $\phi_j\equiv -k\theta \omega_j$,
\begin{align}
    Q\equiv
    \begin{pmatrix}
        2\Gamma^{-1}+\mathbb{1}_M & -U^\text{T}\text{diag}(e^{i\phi_j})_{j=1}^MU^* \\ 
        -U^\dagger\text{diag}(e^{i\phi_j})_{j=1}^MU & 2\Gamma^{-1}+\mathbb{1}_M
    \end{pmatrix},
\end{align}
and, 
\begin{align}
    &\mathcal{N}\equiv \prod_{i=1}^M \frac{\sqrt{1+\gamma_i}}{\pi\gamma_i}, \Gamma\equiv\text{diag}(\gamma_i)_{i=1}^M, \Phi=\text{diag}(e^{i\phi_j}-1)_{j=1}^M, \nonumber \\ 
    &\bm{c}\equiv(\bm{a}^\text{T},\bm{b}^\text{T})^\text{T}, \bm{a}\equiv \frac{U^\text{T}\Phi\bm{\delta}^*}{\sqrt{2}},
    \bm{b}\equiv \frac{U^\dagger\Phi\bm{\delta}}{\sqrt{2}},
    c_0\equiv \frac{\bm{\delta}^\text{T}\Phi\bm{\delta}^*}{2}.
\end{align}
Here, $\bm{\delta}/\sqrt{2}$ is a displacement vector of the final state $\hat{U}|\psi_\text{in}\rangle$.
Note that $Q$ is a complex symmetric matrix and $\text{Re}(Q)$ is positive definite, which guarantees that the Gaussian integral converges.
We provide more details of the derivation and convergence of the integral in Ref.~\cite{supple}.
It implies that molecular vibronic spectra's Fourier components at zero temperature have an analytic solution, so we can obtain the spectra by simply taking the inverse Fourier transform; thus, we do not need a quantum simulator to generate the spectra (see Fig.~\ref{fig:ideal}).
Using the technique from Ref.~\cite{huh2017vibronic}, such an approach can also solve molecular vibronic spectra problem at a finite temperature (see Ref.~\cite{supple} for more details).
In contrast to Fock-state boson sampling case, one interesting feature is that although the spectra is still the sum of probabilities that are hard to compute (hafnian), the spectra itself can be efficiently computed by using the provided method even without approximation.

We emphasize that a similar method that provides an analytical expression for vacuum or thermal input states, corresponding to Gaussian boson sampling, has already been proposed in Ref.~\cite{baiardi2013general}, while our method using positive $P$-representation allows to consider a more general case such as operations mixing position and momentum operators, i.e., a complex unitary operation.
Furthermore, the method in Ref.~\cite{baiardi2013general} overlaps only with Gaussian boson sampling cases and cannot recover more general cases that we consider, such as the Fock-state boson sampling case, which requires a more complicated method.

\begin{table*}
\begin{ruledtabular}
\begin{tabular}{p{0.93in} p{1.15in} p{1.2in} p{1.0in} p{1.4in}}
    & Vacuum or thermal input with squeezing & Fock state & Fock state with squeezing & Fock state with squeezing and displacement \\
    \hline
    \hline
 Related quantity & Gaussian integral* & Permanent* & Hafnian* & Loop hafnian* \\
 \hline
 Complexity & P* & \#P-hard \cite{aaronson2011computational} & \#P-hard \cite{aaronson2011computational, barvinok2016combinatorics}  & \#P-hard \cite{aaronson2011computational,barvinok2016combinatorics} \\
 \hline
 Approximability & Exactly computable* & Approximable* (generalized Gurvits) & Approximable for large squeezing* & Not known \\ 
\end{tabular}
\end{ruledtabular}
\caption{Computational complexity of exactly computing Fourier components of molecular vibronic spectra and approximability within an $1/\poly(n)$ additive error for different setups.
$*$ marks means they are derived in the present work. Specifically, in the present work, we find that the Fourier components of molecular vibronic spectra is related to Gaussian integral, permanent, hafnian, and loop hafnian for different cases, and show that Fock state cases can be solved by a generalized Gurvits's algorithm and Fock state with squeezing by another generalized Gurvits's algorithm (hafnian) for some regime.}
\label{table:exact}
\end{table*}

\subsection*{Potential quantum advantage from molecular vibronic spectra}\label{sec:per}
So far, we have shown that molecular vibronic spectra problem corresponding to Fock-state or Gaussian boson sampling is efficiently solvable using a classical algorithm as accurately as running boson samplers.
The key property we exploit to solve the problem is that the Fourier coefficients are efficiently approximable using generalized Gurvits's algorithm or exactly computable.

We finally consider a more general case, which might potentially provide a quantum advantage.
Here, we set the initial state to be a displaced squeezed Fock state $\hat{D}(\bm{\alpha})\hat{S}(\bm{r}_0)|\bm{n}\rangle$, which can be thought of as a hybrid state of the two previous cases.
We emphasize that such a problem is well-motivated in chemistry in that it is required when a spectra from specific vibronic levels needs to be analyzed, instead of a thermal distribution or a ground state \cite{huh2012coherent}; for example, optical processes including the single vibronic level fluorescence and the resonance Raman scattering \cite{hollas2013high, hollas2004modern}.
Indeed, a quantum simulation of such a process has also been experimentally implemented for various photoelectron processes such as photodetachment of ozone anion \cite{wang2020efficient}.
In this case, its Fourier components are written as
\begin{align}\label{eq:FCP_fock}
    \tilde{G}(k)
    &=\langle\bm{n}|\hat{S}^\dagger(\bm{r}_0)\hat{D}^\dagger(\bm{\alpha})\hat{U}^\dagger e^{-ik\theta \hat{\bm{n}}\cdot\bm{\omega}}\hat{U}\hat{D}(\bm{\alpha})\hat{S}(\bm{r}_0)|\bm{n}\rangle \nonumber \\ 
    &\equiv \langle\bm{n}|\hat{W}|\bm{n}\rangle,
\end{align}
where $\hat{W}\equiv \hat{S}^\dagger(\bm{r}_0)\hat{D}^\dagger(\bm{\alpha})\hat{U}^\dagger e^{-ik\theta \hat{\bm{n}}\cdot\bm{\omega}}\hat{U}\hat{D}(\bm{\alpha})\hat{S}(\bm{r}_0)$ is a general Gaussian unitary operation including squeezing, instead of a linear-optical operation, which is different from the Fock-state boson sampling case.
Using Bloch-Messiah decomposition \cite{braunstein2005squeezing}, we can decompose the Gaussian unitary operation as
$\hat{W}=\hat{D}(\bm{\xi})\hat{U}_\text{lin2}\hat{S}(\bm{r})\hat{U}_\text{lin1}$, where $\hat{U}_{\text{lin1}}$ and $\hat{U}_{\text{lin2}}$ represent linear-optical circuits, which are characterized by $M\times M$ unitary matrices $U_{\text{lin1}}$ and $U_{\text{lin2}}$, respectively.

In Ref.~\cite{supple}, we show that the Fourier components can be written as a loop hafnian of a matrix \cite{bjorklund2019faster}:
\begin{align}\label{eq:loophafnian}
    \tilde{G}(k)=\frac{\lhaf(\tilde{\Sigma}_{\bm{n}})}{\bm{n}!Z},
\end{align}
where
\begin{align}\label{eq:Sigma}
    \Sigma\equiv
    \begin{pmatrix}
        U_\text{lin2}\tanh{\bm{r}}U_\text{lin2}^\text{T} & U_\text{lin2}\sech{\bm{r}}U_\text{lin1} 
        \\ 
        U_\text{lin1}^\text{T}\sech{\bm{r}}U_\text{lin2}^\text{T} & -U_\text{lin1}^\text{T}\tanh{\bm{r}}U_{\text{lin1}}
    \end{pmatrix}
\end{align}
is an $2M\times 2M$ complex symmetric matrix,
\begin{align}
    \bm{\zeta}
    \equiv
    \begin{pmatrix}
        -U_\text{lin2}\tanh{\bm{r}}U_\text{lin2}^\text{T}.\bm{\xi}^*+\bm{\xi} \\ -U_\text{lin1}^\text{T}\sech{\bm{r}}U_\text{lin2}^\text{T}.\bm{\xi}^*
    \end{pmatrix}
\end{align}
is an $2M$-dimensional vector, and 
\begin{align}
    Z^{-1}\equiv \langle 0|\hat{V}|0\rangle
    =\frac{e^{\frac{1}{2}\bm{\xi}^\text{T}.(U_\text{lin2}\tanh{\bm{r}}U_\text{lin2}^\text{T}).\bm{\xi}}}{\sqrt{\prod_{i=1}^M\cosh{r_i}}},
\end{align}
is the normalization factor, which is the same as the Fourier components of molecular vibronic spectra at zero temperature.
Here, $\tilde{\Sigma}_{\bm{n}}$ is obtained by first replacing the diagonal elements of $\Sigma$ by $\bm{\zeta}$ to obtain $\tilde{\Sigma}$ and repeating $i$th row and column of each block matrix of $\tilde{\Sigma}$ $n_i$ times; thus it is an $n\times n$ matrix with $n\equiv\sum_{i=1}^M n_i$.
Therefore, computing the Fourier components reduces to computing loop hafnians of $n\times n$ complex symmetric matrices.

Loop hafnian is a quantity related to perfect matchings of a graph including loops (hafnian does not allow loops.)\cite{bjorklund2019faster}.
The best-known algorithm's computational cost of computing a loop hafnian is $O(n^32^{n/2})$ \cite{bjorklund2019faster} with $n$ being the matrix size.
Thus, if one tries to directly compute the probability of each outcome $\bm{m}$ in Eq.~\eqref{eq:spectra}, which is written as a loop hafnian of a matrix whose size is $n+m$ ($m\equiv \sum_{i=1}^M m_i$) \cite{quesada2019franck}, and then obtain spectra, the complexity of computing a single probability already costs exponential in $(n+m)/2$.
On the other hand, the complexity of computing the Fourier components relies on the input photons $\bm{n}$ not on the output photons $\bm{m}$ and the system size, i.e., the proposed method is efficient as long as $n$ is small enough.
Furthermore, we prove that the redundancy of rows and columns for $n_i\geq 2$ does not increase the complexity of computing a loop hafnian as hafnian \cite{barvinok1996two, bjorklund2019faster, bulmer2021boundary} because it does not increase the rank of the matrix \cite{supple}.
Thus, the important factor for complexity is the number of nonzero elements of $\bm{n}$.

Meanwhile, since loop hafnian is more general than permanent \cite{supple}, computing the loop hafnian is also \#P-hard \cite{aaronson2011computational, barvinok2016combinatorics}.
It implies that computing the Fourier components of general molecular vibronic spectra with Fock state inputs and squeezing is also \#P-hard.
Noting that the Fourier components reduce to hafnian and permanent when there is no displacement or squeezing, respectively,
we summarize the complexity in Table~\ref{table:exact}.

Again, for our purpose, it suffices to find an efficient classical algorithm to additively approximate the Fourier components, which are loop hafnian, to determine if we can efficiently approximate the spectra as accurately as a boson sampler.
In fact, in contrast to permanent, there is no known efficient classical algorithm that approximates loop hafnian as accurately as Gurvtis's algorithm, up to the best of our knowledge.
Thus, one obvious step is to generalize Gurvits's algorithm to be applicable to loop hafnian.
For simplicity, let us assume that the displacement is zero; the loop hafnian in Eq.~\eqref{eq:loophafnian} then reduces to a hafnian.
Similarly to Gurvits's algorithm, we construct a randomized algorithm to estimate a hafnian of an $n\times n$ complex symmetric matrix $\Sigma$ using Kan's formula \cite{kan2008moments}
\begin{align}
    \haf(\Sigma)
    &=\mathbb{E}_{\bm{v}\in\{0,1\}^n}\left[(-1)^{\sum_{i=1}^{n} v_i}\frac{2^{n/2}}{(n/2)!}\left(h^\text{T}\Sigma h\right)^{n/2}\right], \label{eq:kan}
\end{align}
where $\mathbb{E}_{\bm{v}}[\cdot]$ is the average over $\bm{v}\in\{0,1\}^n$ with uniform distribution, and $h=(1/2-v_1,\dots,1/2-v_M)^\text{T}$ and $n$ is even (otherwise the hafnian is zero.).
Therefore, by sampling $\bm{v}\in\{0,1\}^n$ uniformly and averaging over the samples, we estimate the hafnian.
Together with the fact that $\|\Sigma\|=1$ from Eq.~\eqref{eq:Sigma}, the bound of error for estimate $q$ of a Fourier component is given by
\begin{align}
    \text{Pr}\left[\left|q-\frac{\haf(\Sigma_{\bm{n}})}{Z\bm{n}!}\right|\gtrsim \epsilon \frac{e^{n/2}}{\sqrt{\pi n}Z}\right]\leq 2 e^{-N\epsilon^2 /2},
\end{align}
where $Z=\sqrt{\prod_{i=1}^n\cosh{r_i}}$ and $N$ is the number of samples.
Thus, if $\prod_{i=1}^n\cosh{r_i}>e^{n}$, the estimation error is smaller than $\epsilon$ with exponentially small failure probability if we choose the number of samples as $N=O(1/\epsilon^2)$.
Therefore, it suggests that if the condition is satisfied, classically estimating the Fourier components and taking the inverse Fourier transformation efficiently renders the same scaling of precision as a boson sampler.

However, when the condition is not satisfied, the estimation error of the proposed classical algorithm with the generalized Gurvits's algorithm grows exponentially due to the factor $e^{n/2}$.
Therefore, if we cannot find a classical algorithm to approximate hafnian, for example, which approximates such as
\begin{align}
    \text{Pr}\left[\left|q-\frac{\haf(\Sigma_{\bm{n}})}{Z\bm{n}!}\right|\geq \epsilon\right]\leq \delta,
\end{align}
in running time $T=\text{poly}(n,1/\epsilon,\log \delta^{-1})$, a boson sampler might provide a potential advantage for solving this molecular vibronic spectra and give evidence of quantum advantage of boson sampler for practical problems.
If we find such an algorithm, it can eventually solve the problem as accurately as a boson sampler.


For nonzero displacement and loop hafnian, although there is a similar equality as in Eq.~\eqref{eq:kan}, the resultant error bound is more complicated than hafnian (See Ref.~\cite{supple}).
While we did not find a regime where the approximation error is sufficiently small, it would be an important future work to identify the regime to characterize parameters for which running a boson sampling circuit may be advantageous than the classical algorithm.
We finally summarize the well-known or derived complexity of approximation of molecular vibronic spectra's Fourier components in Table.~\ref{table:exact}.

\section*{Discussion}
While we have recently seen the first plausible quantum advantage demonstration experiments using sampling tasks \cite{arute2019quantum, wu2021strong, zhong2020quantum, zhong2021phase, madsen2022quantum}, the present work leaves an open question to find a practically useful task of quantum sampling problems.
Our results imply that while grouping the probabilities and estimating the grouped probability would be a natural and potential way of exploiting boson sampler, an efficient classical counterpart might exist due to the coarse-graining.
In particular, our results suggest that molecular vibronic spectra corresponding to Fock-state or Gaussian boson sampling may not be the candidate for which the power of a quantum sampler can boost the computational performance beyond classical means.
We also note that our method can be easily generalized to certain non-Condon effects \cite{jahangiri2020quantum} and coherent driving of molecular vibronic spectra generation \cite{jnane2021analog}.

On the other hand, we have presented a molecular vibronic spectra problem that might provide a potential quantum advantage and is chemically well-motivated.
For this problem, to the best of our knowledge, the classical algorithm to approximate its Fourier components is not sufficient to perform as accurate as running a boson sampler.
It leaves an interesting open question to find a classical algorithm solving this problem or proving the hardness to identify the potential quantum advantage.
We also note that the way we group the probabilities is based on the simple inner product between the outcome and a given weight vector, which is translated as phase shifters in Fourier basis.
Thus, generalizing the way of grouping the probabilities, which makes the Fourier component to contain a nontrivial non-linear operation, and establishing the complexity would be an important open question to find the power or the limitation of the presented method.

Finally, since we assume harmonic potential for our results, incorporating the anharmonicity of potentials to our results is an important future work.
In Ref.~\cite{supple}, we present a way in which we can embed a BQP-complete problem, which is essentially equivalent to simulating universal quantum computing circuit, to a molecular vibronic spectra problem containing non-linear effects beyond Gaussian operations, which might be necessary to incorporate the anharmonicity.
Thus, the molecular vibronic spectra problem including non-linear effects can be difficult to efficiently simulate using classical computers even in an additive approximation.

\section*{Methods}
\subsection{Molecular vibronic spectra problem}
In this section, we define the molecular vibronic spectra problem from chemical perspective and show its equivalence to the presented definition from boson sampling perspective in the main text.
The molecular vibronic spectra is a fundamental property of molecules, which allows us to extract molecular structural changes.
An electronic transition of a molecule changes the nuclei configuration, which introduces a new set of vibronic modes.
Specifically, the transition probability between an initial vibronic mode and a certain final vibronic mode is called the Franck-Condon (FC) factor.
One may obtain the FC profiles (FCP) by computing many FC factors corresponding to a given vibrational transition frequency ($\Omega$).
The FC factor is obtained with the initial vibrational state $|\psi_\text{in}\rangle$ as
\begin{align}\label{eq:FCP}
    \text{FCP}(\Omega)=\sum_{\bm{m}=\bm{0}}^\infty |\langle \bm{m}|\hat{U}_\text{Dok}|\psi_\text{in}\rangle|^2\delta\left(\Omega-\bm{\omega}^f\cdot\bm{m}\right),
\end{align}
where the Doktorov transformation $\hat{U}_\text{Dok}$ is given by \cite{doktorov1977dynamical}
\begin{align}\label{eq:dok}
    \hat{U}_\text{Dok}=\hat{D}(\bm{\delta}/\sqrt{2})\hat{S}^\dagger(\bm{\Omega}^f)\hat{U}_R\hat{S}(\bm{\Omega}^i),
\end{align}
and $\delta(\cdot)$ is the delta function, and $\bm{m}=(m_1,\dots,m_M)$ is the final vibrational modes' excitation vector.
Here, $\bm{\Omega}^i$ and $\bm{\Omega}^f$ are given by
\begin{align}
    \bm{\Omega}^i\equiv\text{diag}(\sqrt{\omega_1^i},\dots,\sqrt{\omega_M^i}),~
    \bm{\Omega}^f\equiv\text{diag}(\sqrt{\omega_1^f},\dots,\sqrt{\omega_M^f}),
\end{align}
where $\bm{\omega}^i\equiv(\omega_1^i,\dots,\omega_M^i)$ and $\bm{\omega}^f\equiv(\omega_1^f,\dots,\omega_M^f)$ account for initial and final harmonic angular frequencies, respectively.

In particular, the FC factor at zero temperature is obtained with the initial vacuum state $|\psi_\text{in}\rangle=|\bm{0}\rangle$.
In this case, using the Bloch-Messiah decomposition, we can rewrite the Doktorov transformation as
\begin{align}
    \hat{U}_\text{Dok}
    =\hat{U}_2 \hat{D}(\bm{\alpha})\hat{S}(\bm{\Omega}'')\hat{U}_1,
\end{align}
where $\hat{U}_1$ and $\hat{U}_2$ are rotations, i.e., described by linear-optical circuits.
Since $\hat{U}_1|\bm{0}\rangle=|\bm{0}\rangle$, the final state after the Doktorov transformation is given by
\begin{align}
    \hat{U}_\text{Dok}|\bm{0}\rangle
    =\hat{U}_2 \hat{D}(\bm{\alpha})\hat{S}(\bm{\Omega}'')|\bm{0}\rangle,
\end{align}
which is exactly the same as the output state of Gaussian boson sampling with arbitrary pure product Gaussian states.
It shows the direct relation between the molecular vibronic spectra at zero temperature and Gaussian boson sampling.

In the main text, we also study the molecular vibronic spectra from a single vibronic level, where the input state is no longer a vacuum state but a Fock state $|\bm{n}\rangle$.
In this case, the output state is given by $\hat{U}_\text{Dok}|\bm{n}\rangle$.
Especially when we consider the molecular vibronic spectra problem corresponding to Fock-state boson sampling, the Doktorov transformation is simply a rotation without any squeezing and displacement.

Let us discuss the relation between the frequency vector and the weight vector $\bm{\omega}$ further.
First of all, since we only consider a Fock state or a vacuum state input, the initial energy is always fixed; thus, it is merely an offset, which we can omit.
Hence, the relation is between the final frequency vector $\bm{\omega}^f$ and $\bm{\omega}$.
Although the frequency vector $\bm{\omega}$ does not necessarily an integer vector, the resolution of $\bm{\omega}^f$ is limited in practical molecular vibronic spectra problems.
Hence, practically, the frequency vector can be written as floating point numbers up to the resolution.
Therefore, by multiplying a sufficiently large number, we can transform the actual frequency into a weight vector which is an integer vector.
Here, the magnitude of the final weight vector depends on the resolution of the frequencies.
For example, if we have a polynomial resolution on the frequencies, we only need to multiply a polynomially large number to transform the vector to an integer weight vector.
If we have an exponential resolution on the frequencies, we need to multiply an exponentially large number.
Therefore, assuming the weight vector to be an integer does not lose generality.


We note that throughout the work, we assume harmonic potential for vibrational modes while we emphasize that the effect of anharmonicity is important for more general cases \cite{mcardle2019digital, sawaya2019quantum, sawaya2021near}.

\subsection{Derivation of Fourier coefficient}
Here, we derive the expression of Fourier coefficients $\tilde{G}(k)$ of the spectra $G(\Omega)$ in Eq.~\eqref{eq:phase}.
\begin{align}
    \tilde{G}(k)
    &\equiv \sum_{\Omega=0}^{\Omega_\text{max}}G(\Omega)e^{-ik\theta\Omega} \\ 
    &=\sum_{\Omega=0}^{\Omega_\text{max}}\sum_{\bm{m}=0}^\infty p(\bm{m})\delta(\Omega-\bm{\omega}\cdot \bm{m})e^{-ik\theta \Omega} \\
    &=\frac{1}{\Omega_\text{max}+1}\sum_{\Omega=0}^{\Omega_\text{max}}\sum_{\bm{m}=0}^\infty p(\bm{m})\sum_{l=0}^{\Omega_\text{max}}e^{il\theta(\Omega-\bm{\omega}\cdot \bm{m})-ik\theta \Omega} \\
    &=\frac{1}{\Omega_\text{max}+1}\sum_{l=0}^{\Omega_\text{max}}\sum_{\bm{m}=0}^\infty p(\bm{m})e^{-il\theta\bm{\omega}\cdot \bm{m}}\sum_{\Omega=0}^{\Omega_\text{max}}e^{i\theta\Omega(l-k)} \\ 
    &=\sum_{\bm{m}=0}^\infty p(\bm{m})e^{-ik\theta \bm{\omega}\cdot\bm{m}} \\
    &=\sum_{\bm{m}=0}^\infty \langle \psi_\text{out}|\bm{m}\rangle\langle \bm{m}|\psi_\text{out}\rangle e^{-ik\theta \bm{\omega}\cdot\bm{m}} \\&=\left\langle \bigotimes_{i=1}^M \left(\sum_{m_i=0}^\infty |m_i\rangle\langle m_i|e^{-ik\theta \omega_im_i}\right)\right\rangle \\ 
    &=\langle e^{-ik\theta  \bm{\omega}\cdot\hat{\bm{n}}}\rangle.
\end{align}
Also, the Fourier relation also implies that
\begin{align}
    G(\Omega)=\frac{1}{\Omega_\text{max}+1}\sum_{k=0}^{\Omega_\text{max}} \tilde{G}(k)e^{ik\theta\Omega},
\end{align}
which is Eq.~\eqref{eq:G}.

\subsection{Error propagation from inverse Fourier transformation}
We present the relation between the approximation error in Fourier basis and the one in spectra using Parseval relation.
Let us assume that we have obtained the Fourier components $\tilde{G}(k)$ within an error $\epsilon$, i.e., $|\Delta \tilde{G}(k)|\leq \epsilon$.
After taking the inverse Fourier transform $\mathcal{F}^{-1}$, we obtain $G(\Omega)$ with an error $\Delta G(\Omega)$:
\begin{align}
    G(\Omega)+\Delta G(\Omega)=\mathcal{F}^{-1}[ \tilde{G}(k)+\Delta \tilde{G}(k)],
\end{align}
and
\begin{align}
    |\Delta G(\Omega)|=|\mathcal{F}^{-1}[\Delta \tilde{G}(k)]|.
\end{align}
Using the Parseval relation, we have
\begin{align}\label{eq:parseval}
    \sum_{\Omega}|\Delta G(\Omega)|^2=\frac{1}{K+1}\sum_{k}|\Delta \tilde{G}(k)|^2\leq \epsilon^2,
\end{align}
which shows $|\Delta G(\Omega)|\leq \epsilon$ for all $\Omega$.
Therefore, if we estimate the Fourier components within an error $\epsilon$ for each, then the error for the spectra is smaller than $\epsilon$.

\subsection{Sparse fast Fourier transformation}
As stated in the main text, we exploit the sparse fast Fourier transformation~\cite{hassanieh2012nearly} for the case where the weight vectors are superpolynomially large, so that the total number of bins is $\Omega_\text{max}+1=\omega(\poly(n))$ and show how it enables us to classically solve the problem efficiently.

The sparse fast Fourier transformation runs as follows (more details can be found in Ref.~\cite{hassanieh2012nearly}):
(1) By randomly permuting the bins and hashing the bins, obtain a smaller number of bins, each of which contains at most a single peak from the initial distribution with high probability.
(2) Locate the peaks from the hashed bins.
(3) Estimate a large constant fraction of the peaks with good precision with high probability.
(4) Subtract the obtained estimates, which makes the distribution sparser.
(5) Iterate the procedure (1)-(4).
The algorithm takes the running time $O(t/\epsilon \log (\Omega_\text{max}/t)\log (\Omega_\text{max}/\delta))$ time and succeeds with high probability, where $\epsilon$ and $\delta$ determine the error.

The sparse fast Fourier transformation is already sufficient to solve the molecular vibronic spectra problem corresponding to Gaussian boson sampling.
For the problem corresponding to Fock-state boson sampling, we need to be more careful because we can only approximate the Fourier coefficients.
In Ref.~\cite{supple}, we show that the approximation error does not change the position of peaks, i.e., it does not delete or produce peaks, with high probability, which guarantees that the sparse fast Fourier transformation with the generalized Gurvits's algorithm solves the problem efficiently.

\subsection{Generalized and positive $P$-representation}
The generalized $P$-representation of an $M$-mode quantum optical state is one of the quasi-probability distributions of a bosonic state \cite{drummond1980generalised, walls2007quantum, supple}:
\begin{align}\label{eq:p_def}
    \hat{\rho}=
    \int_{\mathbb{C}^{2M}} P(\bm{\alpha},\bm{\beta})\hat{\Lambda}(\bm{\alpha},\bm{\beta})d\bm{\alpha}d\bm{\beta},
    ~~~
    \hat{\Lambda}(\bm{\alpha},\bm{\beta})\equiv \frac{|\bm{\alpha}\rangle\langle \bm{\beta}^*|}{\langle \bm{\beta}^*|\bm{\alpha}\rangle},
\end{align}
where $|\bm{\alpha}\rangle=|\alpha_1\rangle\otimes\dots\otimes |\alpha_M\rangle$ is an $M$-mode coherent state.
An important property of the generalized $P$-representation is that the distribution $P(\bm{\alpha},\bm{\beta})$ can always be chosen nonnegative (we call this positive $P$-representation), and the expectation value of a normal-ordered operator can be readily computed \cite{drummond1980generalised}.
Especially, the positive $P$-representation for a single-mode squeezed state can be chosen as \cite{drummond1980generalised, janszky1990squeezing, drummond2021simulating, supple}
\begin{align}\label{eq:p_sq}
    P(x,y)=\frac{\sqrt{1+\gamma}}{\pi \gamma} e^{-(x^2+y^2)(\gamma^{-1}+1/2)+xy},
\end{align}
where $x$ and $y$ are real numbers (not complex numbers as general cases in Eq.~\eqref{eq:p_def}) and $\gamma\equiv e^{2r}-1$ for a squeezing parameter $r>0$.

\begin{acknowledgements}
We thank Nicolás Quesada, Joonsuk Huh, Sandy Irani, Bryan O'Gorman, James Whitfield, and Nicolas Sawaya for interesting and fruitful discussions.
B.F. acknowledges support from AFOSR (YIP number FA9550-18-1-0148 and FA9550-21-1-0008). This material is based upon work partially supported by the National Science Foundation under Grant CCF-2044923 (CAREER) and by the U.S. Department of Energy, Office of Science, National Quantum Information Science Research Centers as well as by DOE QuantISED grant DE-SC0020360.
We acknowledge support from the ARO (W911NF-18-1-0020, W911NF-18-1-0212), ARO MURI (W911NF-16-1-0349, W911NF-21-1-0325), AFOSR MURI (FA9550-19-1-0399, FA9550-21-1-0209), AFRL (FA8649-21-P-0781), DoE Q-NEXT, NSF (EFMA-1640959, OMA-1936118, EEC-1941583, OMA-2137642), NTT Research, and the Packard Foundation (2020-71479).
This research was supported in part by the National Science Foundation under PHY-1748958.
Y. L. acknowledges National Research Foundation of Korea a grant funded by the Ministry of Science and ICT (NRF-2020M3E4A1077861) and KIAS Individual Grant
(CG073301) at Korea Institute for Advanced Study.
We also acknowledge the University of Chicago’s Research Computing Center for their support of this work.
\end{acknowledgements}

\bibliography{reference.bib}

\end{document}


\definecolor{navy}{RGB}{46,72,102}
\definecolor{pink}{RGB}{219,48,122}
\definecolor{grey}{RGB}{184,184,184}
\definecolor{yellow}{RGB}{255,192,0}
\definecolor{grey1}{RGB}{217,217,217}
\definecolor{grey2}{RGB}{166,166,166}
\definecolor{grey3}{RGB}{89,89,89}
\definecolor{red}{RGB}{255,0,0}

\preprint{APS/123-QED}

\title{Supplemental Material for ``Quantum-inspired classical algorithm for molecular vibronic spectra"}
\author{Changhun Oh}
\email{changhun@uchicago.edu}
\affiliation{Pritzker School of Molecular Engineering, University of Chicago, Chicago, Illinois 60637, USA}
\author{Youngrong Lim}%
\affiliation{School of Computational Sciences, Korea Institute for Advanced Study, Seoul 02455, Korea}
\author{Yat Wong}
\affiliation{Pritzker School of Molecular Engineering, University of Chicago, Chicago, Illinois 60637, USA}
\author{Bill Fefferman}
\affiliation{Department of Computer Science, University of Chicago, Chicago, Illinois 60637, USA}
\author{Liang Jiang}
\affiliation{Pritzker School of Molecular Engineering, University of Chicago, Chicago, Illinois 60637, USA}

\maketitle

\section{Gurvits's algorithm with repeated rows and columns}\label{sec:GGurvits}
We first present the original Gurvits's algorithm and generalize the algorithm to improve the bound when rows and columns are repeated.
Gurvits's algorithm exploits the following equality \cite{gurvits2005complexity, aaronson2011computational}:
\begin{align}
    \per(X)=\mathbb{E}_{\bm{x}\in\{-1,1\}^n}\left[\prod_{i=1}^n \left(\sum_{j=1}^n x_iX_{ij}x_j\right)\right].
\end{align}
Thus, by sampling the random variable with uniform $\bm{x}\in\{-1,1\}^n$, the randomized algorithm gives an estimate $\mu$ of permanent of an $n\times n$ matrix $X$ such that
\begin{align}
    |\per(X)-\mu|<\epsilon\|X\|^n,
\end{align}
with high probability $1-\delta$ with running time $T=O(\text{poly}(n,1/\epsilon,\log\delta^{-1})$).
When we consider a submatrix of a unitary matrix, this expression is sufficient because the submatrix's norm is bounded by 1.
However, when we have repeated rows and columns in $X$ from a unitary matrix, which is the case when we have multiphoton inputs, the norm $\|X\|$ is not generally bounded by 1.
Therefore, the upper bound of the error can increase as $\|X\|^n$, i.e., exponentially in $n$.
To deal with multiphoton cases with a reduced error, we need to generalize Gurvits's algorithm.
Note that although Ref.~\cite{aaronson2012generalizing} generalizes Gurvits's algorithm for multiphoton outputs, i.e., $\per(V_{\bm{n},\bm{m}})=\langle\bm{n}|\hat{V}|\bm{m}\rangle$, where $\bm{n}\in \{0,1\}^n$ and $\bm{m}$ is a nonnegative integer vector, it is not sufficient for our purpose where we have multiphoton input and output, i.e., $\langle \bm{n}|\hat{V}|\bm{n}\rangle=\per(V_{\bm{n},\bm{n}})$.

Consider a case where the goal is to approximate a quantity
\begin{align}
    \frac{\per(A)}{\bm{n}!},
\end{align}
where $A\in \mathbb{C}^{n\times n}$ which is obtained by repeating $i$th row and column of a matrix $B\in \mathbb{C}^{k\times k}$ for $n_i$ times.
Here, $\bm{n}\in\mathbb{Z}_{\geq 0}^k$ with $\sum_{i=1}^k n_i=n$.
Following the method in Ref.~\cite{aaronson2012generalizing}, let us define a random variable
\begin{align}
    \text{GenGly}_x(A)\equiv \frac{1}{\prod_{i=1}^{k}n_i^{n_i}}\prod_{i=1}^k\bar{y}_i^{n_i}\prod_{i=1}^k(y_1B_{i,1}+\cdots+y_kB_{i,k})^{n_i}
    =\prod_{i=1}^k\left(\frac{\bar{y}_i (By)_i}{n_i}\right)^{n_i},
\end{align}
where $y_i\equiv n_i^{1/2}x_i$.
Here, $\bm{x}\in \mathcal{R}[n_1+1]\times \cdots \times \mathcal{R}[n_k+1]\equiv \mathcal{X}$, where $\mathcal{R}[j]$ is the set of $j$th roots of unity.
Then, its absolute value is given as
\begin{align}
    |\text{GenGly}_x(A)|
    =\left(\prod_{i=1}^{k}n_i^{n_i}\right)^{-1/2}\left|\prod_{i=1}^k(y_1b_{i,1}+\cdots+y_kb_{i,k})^{n_i}\right|
    =\left(\prod_{i=1}^{k}n_i^{n_i}\right)^{-1/2}\left|\prod_{i=1}^kc_i^{n_i}\right|,
\end{align}
where $c_i\equiv y_1B_{i,1}+\cdots+y_kB_{i,k}$.
Now, let us find its upper bound, which determines the error of the corresponding randomized algorithm by Chernoff bound.
First, we have a constraint
\begin{align}\label{eq:constraint}
    \sum_{i=1}^k |c_i|^2
    =\sum_{i=1}^k |(By)_i|^2
    \leq \|B\|^2 \|y\|^2
    =n\|B\|^2.
\end{align}
We then consider
\begin{align}
    \log \prod_{i=1}^k |c_i|^{n_i}
    =\sum_{i=1}^k n_i \log |c_i|.
\end{align}
Using lagrange multiplier, with a constraint $\sum_{i=1}^k|c_i|^2=c$,
\begin{align}
    Z\equiv \sum_{i=1}^k n_i \log |c_i|-\lambda\left(\sum_{i=1}^k |c_i|^2-c\right),
\end{align}
we obtain
\begin{align}
    \frac{\partial Z}{\partial |c_i|}=\frac{n_i}{|c_i|}-2\lambda |c_i|=0, ~~\forall i.
\end{align}
Using the constraint Eq.~\eqref{eq:constraint},
\begin{align}
    \sum_{i=1}^k|c_i|^2=\sum_{i=1}^k\frac{n_i}{2\lambda}
    =\frac{n}{2\lambda}=c,
\end{align}
the solution is obtained as
\begin{align}
    |c_i|=\sqrt{\frac{n_i}{2\lambda}}
    =\sqrt{\frac{cn_i}{n}}.
\end{align}
Therefore, using the solution, we have an upper bound as
\begin{align}
    \prod_{i=1}^k|c_i|^{n_i}
    \leq \exp\left(\sum_{i=1}^k n_i\log \sqrt{\frac{cn_i}{n}}\right)
    =\prod_{i=1}^k n_i^{n_i/2}\left(\frac{c}{n}\right)^{n/2}
    \leq \prod_{i=1}^k n_i^{n_i/2} \|B\|^n.
\end{align}
Hence, we have an upper bound of the absolute value of the random variable,
\begin{align}
    |\text{GenGly}_x(A)|=\left(\prod_{i=1}^{k}n_i^{n_i}\right)^{-1/2}\left|\prod_{i=1}^kc_i^{n_i}\right|
    \leq \|B\|^n.
\end{align}

Now, we prove that the average of the random variables gives an unbiased estimator of $\per(A)/\bm{n}!$, i.e.,
\begin{align}
    \mathbb{E}_{\mathbb{x}\in \mathcal{X}}[\text{GenGly}_x(A)]
    =\frac{\per(A)}{\bm{n}!}.
\end{align}
First, we have
\begin{align}
    \mathbb{E}_{\mathbb{x}\in \mathcal{X}}
    \left[\frac{1}{\prod_{i=1}^{k}n_i^{n_i}}\prod_{i=1}^k\bar{y}_i^{n_i}\prod_{i=1}^k(y_1B_{i,1}+\cdots+y_kB_{i,k})^{n_i}\right]
    &=\frac{1}{\prod_{i=1}^{k}n_i^{n_i}}\mathbb{E}_{ \mathbb{x}\in \mathcal{X}}
    \left[\prod_{i=1}^k\bar{y}_i^{n_i}(y_1B_{i,1}+\cdots+y_kB_{i,k})^{n_i}\right] \\ 
    &=\frac{1}{\prod_{i=1}^{k}n_i^{n_i}}
    \sum_{\sigma_1,\dots,\sigma_n \in[k]}\left(\prod_{i=1}^k B_{i,\sigma_i}\right)
    \mathbb{E}_{ \mathbb{x}\in \mathcal{X}}
    \left[\prod_{i=1}^k\bar{y}_i^{n_i} \prod_{i=1}^n y_{\sigma_i}\right],
\end{align}
where by the symmetry over the roots of unity,
\begin{align}
    \mathbb{E}_{ \mathbb{x}\in \mathcal{X}}
    \left[\prod_{i=1}^k\bar{y}_i^{n_i} \prod_{i=1}^n y_{\sigma_i}\right]=0,
\end{align}
unless the product insides is equal to $\prod_{i=1}^k|y_i|^{2n_i}$,
in which case we have
\begin{align}
    \mathbb{E}_{ \mathbb{x}\in \mathcal{X}}
    \left[\prod_{i=1}^k\bar{y}_i^{n_i} \prod_{i=1}^n y_{\sigma_i}\right]
    =\prod_{i=1}^k n_i^{n_i}.
\end{align}
Hence,
\begin{align}
    \mathbb{E}_{\mathbb{x}\in \mathcal{X}}[\text{GenGly}_x(A)]
    =\sum_{t_1\dots,t_k\in [k]:|\{i:t_i=j\}|=n_j~\forall j\in[k]}\prod_{i=1}^k B_{i,t_i}
    =\frac{\per(A)}{\bm{n}!},
\end{align}
where $\bm{n}!$ appears to take into account the fact that each produce $\prod_{i=1}^k b_{i,t_k}$ appears $\bm{n}!$ times.

Therefore, $\text{GenGly}_x(A)$ gives us an unbiased estimator of $\per(A)/\bm{n}!$, and the random variables are bounded by $\|B\|^n$.
Due to the Chernoff bound, the randomized algorithm with number of samples $N=O(1/\epsilon^2)$ provides an estimate $\mu$ with high probability $1-\delta$ such that
\begin{align}
    \left|\frac{\per(A)}{\bm{n}!}-\mu\right|< \epsilon\|B\|^n.
\end{align}

\section{Sparse fast Fourier transformation for large weight vector}
In the main text, we showed that we can efficiently solve the problem where $\Omega_\text{max}=O(\poly(M))$ due to the assumption that $\omega_i=O(\poly(M))$ for all $i$'s.
By lifting the assumption, we now consider the case where $\Omega_\text{max}=\omega(\poly(M))$.
Let $d\equiv \Omega_\text{max}+1$ be the number of bins.
In particular, we will assume that the goal is reproducing the $t=O(\poly(M))$ largest components $G(\Omega)$ in the spectrum, namely peaks, instead of reproducing all the components to focus on the problems that a boson sampler can solve in polynomial time.
This is because estimating a superpolynomial number of quantities inevitably takes superpolynomial time.
This assumption is also practically reasonable because the molecular vibronic spectra problem's typical purpose is identifying the peaks.
For a boson sampler to achieve reasonable accuracy, we assume that the $t$ peaks are as large as $1/\poly(M)$.
We further assume that $\log \omega_i=O(\poly(M))$ for all $i\in [M]$ so that $\omega_i$'s can be represented in a polynomial number of bits.
Consequently, $\log d=O(\poly(M))$.



\subsection{Boson-sampling case}
Suppose that we have obtained $N$ number of samples by running a boson sampler.
The standard Chernoff bound states that for each $\Omega\in [0,\Omega_\text{max}]$ with the $N$ samples, we obtain the estimate of $G(\Omega)$ with accuracy $\epsilon$ with high probability $1-2e^{-2N\epsilon^2}$.
By applying the union bound for all the $t$ peaks, we obtain the accuracy $\epsilon$ for $t$ different $\Omega$ with probability $1-2te^{-2N\epsilon^2}=1-2e^{-2N\epsilon^2+\log t}$.
Thus, we choose $N$ such that $-2N\epsilon^2+\log t=O(1)$.
Then, we need at least $N=O(\log t/\epsilon^2)$.
Since our goal is to estimate $G(\Omega)$ for all the $t$ peaks, we can guarantee that the estimated peaks $G_t'$ is $t\epsilon$-close to the true peaks:
\begin{align}
    \|G_t-G_t'\|_1\leq t\epsilon,
\end{align}
where $G_t$ is the distribution only for the peaks.
In other words, $N=O(t^2\log t/\epsilon^2)$ is sufficient to guarantee with high probability that 
\begin{align}
    \|G_t-G_t'\|_1\leq \epsilon.
\end{align}

\subsection{Efficient classical algorithm using a sparse fast Fourier transformation}
We now consider the same problem, i.e., estimating the $t$ peaks $G(\Omega)=\Omega(1/\poly(M))$.
For our classical algorithm, we exploit the sparse fast Fourier transformation (SFFT) \cite{hassanieh2012nearly}.
For general cases, the SFFT gives the estimate $G'=G'(\Omega)$ such that
\begin{align}
    \|G-G'\|_2
    \leq (1+\epsilon)\min_{H\in t\text{-sparse}}\|G-H\|_2+\delta \|G\|_2,
\end{align}
in $O(t/\epsilon \log (d/t)\log (d/\delta))$ time with high probability.
Here, the minimization is over $t$-sparse distributions $H=H(\Omega)$, i.e., only $t$ elements are non-zero.
For our case, $\|G\|_2\leq \|G\|_1=1$, and due to the dependence of the running time on $\delta$, we can choose $\delta=1/\exp(M)$.
Intuitively, the multiplicative factor $1$ comes from the tails, i.e., the terms that are not peaks; thus, if we focus only on the peaks, the right-hand-side should be efficiently reduced by choosing $\epsilon$ and $\delta$ to be $1/\poly(M)$.

More rigorously, we can show that the error on the peaks can be efficiently reduced by the fact that the final round of the algorithm in Ref.~\cite{hassanieh2012nearly} gives us the estimate $G'(\Omega)$ of the distribution $G(\Omega)$ (see Eq.~(6) in Ref.~\cite{hassanieh2012nearly}), which satisfies
\begin{align}
    \|G-G'\|_\infty^2\leq \frac{\epsilon}{t}\left(\min_{H\in t\text{-sparse}}\|G-H\|_2^2+d\delta^2\right).
\end{align}
Therefore, the total variation distance for the peaks is given by
\begin{align}
    \|G_t-G_t'\|_1\leq \sqrt{t\epsilon}\left(\min_{H\in t\text{-sparse}}\|G-H\|_2^2+d\delta^2\right)^{1/2}.
\end{align}
Hence, choosing $\epsilon'=\sqrt{t\epsilon}\left(\min_{H\in t\text{-sparse}}\|G-H\|_2^2+d\delta^2\right)^{1/2}$ with $\delta=1/\sqrt{d}$, we can achieve 
\begin{align}
    \|G_t-G_t'\|_1\leq \epsilon',
\end{align}
in $O(t^2/\epsilon'^2 \log (d/t)\log (d/\delta))$ time with high probability.
Hence, when we can compute the Fourier components $\tilde{G}(k)$ exactly, the SFFT gives an estimate $G'$ with high probability.
Thus, it already suffices to show that for the molecular vibronic spectra problem corresponding to the Gaussian boson sampling, the SFFT provides an estimate of $G(\Omega)$ as accurate as running a Gaussian boson sampler.

For the case of Fock-state boson sampling, we can only approximate the Fourier components because exactly computing them is \#P-hard.
Therefore, the SFFT has to be applied to the approximated Fourier components whose inverse Fourier transformation gives $\hat{G}$.
If we assume that the approximation of the Fourier components does not change the position of peaks (it may change the value of the peaks but not the position of the peaks), the SFFT applied to the approximated Fourier coefficients can still recover the peaks with error produced by the approximation.
Thus, under the assumption, we have the following inequality for the estimated spectrum $G_t'$:
\begin{align}
    \|G_t-G_t'\|_1
    \leq \|G_t-\hat{G}_t\|_1+\|\hat{G}_t-G_t'\|_1
    \leq \sqrt{t} \epsilon''+\epsilon',
\end{align}
where $\epsilon'$ comes from the SFFT, $\epsilon''$ comes from the approximation of the Fourier components (See Methods C.), and $\sqrt{t}$ factor comes from $\|v\|_1\leq \sqrt{t}\|v\|_2$ for length $t$ vector $v$.
Thus, by setting $\epsilon''=\epsilon/(2\sqrt{t})$ and $\epsilon'=\epsilon/2$, we obtain the accuracy $\|G_t-G_t'\|_1\leq \epsilon$ in running time $O(\poly(M, 1/\epsilon))$.

Now, the remaining problem is to show that the approximation of the Fourier coefficients does not change the position of the peaks.
To prove this, it suffices to show that the approximation error in the Fourier basis does not generate an error as large as $o(1/\poly(M))$ with high probability.
To this end, we make use of the Hoeffding inequality for the sub-Gaussian distribution.
This is because we approximate the permanent using the generalized Gurvits's algorithm.
In particular, the Hoeffding inequality for the generalized Gurvits's algorithm from Sec.~\ref{sec:GGurvits} gives the following inequality:
\begin{align}
    \text{Pr}\left[\left|\tilde{p}-\frac{\per(V_{\bm{n}})}{\bm{n}!}\right|\geq \epsilon\right]\leq 2\exp\left(-\frac{\epsilon^2 N}{2}\right),
\end{align}
where $N$ is the number of samples.
On the other hand, the definition of the sub-Gaussian distribution is that
\begin{align}
    \text{Pr}\left[|X|\geq t\right]\leq 2e^{-ct^2},
\end{align}
for some $c>0$.
Thus, the estimation error from the generalized Gurvits's algorithm follows a sub-Gaussian distribution for any $N>0$.

For random sub-Gaussian variables $X_1,\dots,X_m$, the Hoeffiding inequality gives \cite{vershynin2018high}
\begin{align}\label{eq:hoeffding_subG}
    \text{Pr}_X\left[\left|\frac{\sum_{i=1}^m X_i}{m}\right|\geq \epsilon\right]
    \leq 2\exp\left(-\frac{cm^2\epsilon^2}{\sum_{i=1}^m\|X_i\|_{\psi_2}}\right)
    \equiv 2\exp\left(-Cm\epsilon^2\right)
\end{align}
for an absolute constant $c>0$, and $C=cm/\sum_{i=1}^m \|X_i\|_{\psi_2}>0$.
Here, we defined
\begin{align}
    \|X_i\|_{\psi_2}\equiv \inf\{c>0:\mathbb{E}[e^{X_i^2/c^2}]\leq 2\},
\end{align}
which is finite if and only if $X_i$ is sub-Gaussian.
Now, recall that the error of the spectra is written as
\begin{align}
    \Delta G(\Omega)=\frac{1}{\Omega_\text{max}+1}\sum_{k=0}^{\Omega_\text{max}}\Delta\tilde{G}(k)e^{ik\theta\Omega}.
\end{align}
The real part of the error is written as
\begin{align}\label{eq:real_error}
    \Re[\Delta G(\Omega)]
    =\frac{1}{\Omega_\text{max}+1}\sum_{k=0}^{\Omega_\text{max}}\Re[\Delta \tilde{G}(k)e^{ik\theta \Omega}]
    =\frac{1}{\Omega_\text{max}+1}\sum_{k=0}^{\Omega_\text{max}}\left[\Re[\Delta \tilde{G}(k)]\cos (k\theta \Omega)-\Im[\Delta \tilde{G}(k)]\sin (k\theta \Omega)\right].
\end{align}
To avoid the possibility that the real and imaginary errors are correlated, we assume that the real part and the imaginary part are estimated independently in the generalized Gurvits's algorithm, which only doubles the computational cost.
By applying the Hoeffding bound \eqref{eq:hoeffding_subG} on Eq.~\eqref{eq:real_error}, we obtain
\begin{align}
    \text{Pr}\left[\left|\text{Re}[\Delta G(\Omega)]\right|\geq \epsilon\right]\leq 2\exp\left(-C_Rd\epsilon^2\right),
\end{align}
where $C_R>0$ is a constant.
Using the same argument for the imaginary part and applying the union bound, we obtain
\begin{align}
    \text{Pr}\left[\left|\Delta G(\Omega)\right|\geq \epsilon\right]\leq 2\exp\left(-C_Rd\epsilon^2\right)+2\exp\left(-C_Id\epsilon^2\right)
    \leq 4\exp\left(-Cd\epsilon^2\right),
\end{align}
where $C_I$ is the counterpart of $C_R$ for imaginary part and $C\equiv \min(C_R,C_I)$.
Now we apply the union bound for all $\Omega$, which gives us
\begin{align}
    \text{Pr}\left[\left|\Delta G(\Omega)\right|\geq \epsilon~\text{for some}~ \Omega\in [0, \Omega_\text{max}]\right]\leq 4d\exp\left(-Cd\epsilon^2\right)
    =4\exp\left(-Cd\epsilon^2+\log d\right).
\end{align}
Here, by choosing $\epsilon=1/d^{1/4}=o(1/\poly(M))$, we finally prove that the approximation error in the spectrum is upper bounded by $o(1/\poly(M))$ with high probability.

\section{The positive $P$-representation}\label{sec:P}
In this section, we recall the generalized (positive) $P$-representation introduced in Ref.~\cite{drummond1980generalised}:
\begin{align}
    \hat{\rho}=
    \int_{\mathbb{C}^{2M}} P(\bm{\alpha},\bm{\beta})\hat{\Lambda}(\bm{\alpha},\bm{\beta})d\bm{\alpha}d\bm{\beta},
    ~~~
    \hat{\Lambda}(\bm{\alpha},\bm{\beta})\equiv \frac{|\bm{\alpha}\rangle\langle \bm{\beta}^*|}{\langle \bm{\beta}^*|\bm{\alpha}\rangle},
\end{align}
where $|\alpha\rangle$ and $|\beta^*\rangle$ are coherent states.
The generalized $P$-representation is one of the quasi-probability distributions of a bosonic state \cite{walls2007quantum}.
An important property of the generalized $P$-representation is that the distribution $P(\bm{\alpha},\bm{\beta})$ can always be chosen nonnegative (we call this positive $P$-representation.) and the expectation value of a normal-ordered operator can be readily computed \cite{drummond1980generalised}.
The former can be proved as follows.
For simplicity, we use a simpler version, where we assume that the Glauber-Sudarshan $P$-function exists \cite{walls2007quantum}, which is written as
\begin{align}\label{eq:q}
    \hat{\rho}=\int_{\mathbb{C}}d^2\alpha P_{GS}(\alpha)|\alpha\rangle\langle \alpha|,
\end{align}
for a single-mode state.
We note that although we assume the existence of the Glauber-Sudarshan $P$-function for simplicity, the claim is true for arbitrary density matrices \cite{drummond1980generalised}.
The claim is that the positive $P$-representation is written as
\begin{align}
    P(\alpha,\beta)=\frac{1}{4\pi^2}\exp\left(-\frac{|\alpha-\beta^*|^2}{4}\right)\langle(\alpha+\beta^*)/2|\hat{\rho}|(\alpha+\beta^*)/2\rangle.
\end{align}
If the claim is true, then it guarantees that the representation is positive and real.
By directly substituting the representation, we get
\begin{align}\label{eq:p}
    P(\alpha,\beta)=\frac{1}{4\pi^2}\int P_\text{GS}(\gamma)\exp\left(-\frac{1}{2}|\alpha-\gamma|^2-\frac{1}{2}|\beta^*-\gamma|^2\right)d^2\gamma.
\end{align}
Meanwhile, for an analytical function $f$, we have
\begin{align}
    f(\gamma)=\frac{1}{2\pi}\int_{\mathbb{C}} f(\alpha)\exp\left(-\frac{1}{2}|\alpha-\gamma|^2\right)d^2\alpha.
\end{align}
It can be shown that for example at $\gamma=0$,
\begin{align}
    \int_{\mathbb{C}} f(\alpha)\exp\left(-\frac{1}{2}|\alpha|^2\right)d^2\alpha
    =2\pi \exp\left(\frac{1}{2}\frac{\partial^2}{\partial \alpha \partial \alpha^*}\right)f(\alpha)|_{\alpha=0}=2\pi f(0),
\end{align}
where the last equality holds because the function is analytic.
Using this property, we obtain from Eq.~\eqref{eq:p}
\begin{align}
    \int_{\mathbb{C}^2} \hat{\Lambda}(\alpha,\beta)P(\alpha,\beta)d^2\alpha d^2\beta
    &=\frac{1}{4\pi^2}\int_{\mathbb{C}^3} d^2\alpha d^2\beta d^2\gamma \hat{\Lambda}(\alpha,\beta)P_\text{GS}(\gamma)\exp\left(-\frac{1}{2}|\alpha-\gamma|^2-\frac{1}{2}|\beta^*-\gamma|^2\right) \\ 
    &=\int_{\mathbb{C}} d^2\alpha P_\text{GS}(\alpha)|\alpha\rangle\langle \alpha|=\hat{\rho},
\end{align}
which proves the claim.

Now, we show by an example that the expectation value of a normal-ordered operator can be computed by averaging over the phase-space variable.
For example, when one computes for a single-mode $:\hat{a}\hat{a}^\dagger\hat{a}\hat{a}^\dagger:=\hat{a}^{\dagger2}\hat{a}^2$,
\begin{align}
    \Tr[\hat{\rho}:\hat{a}\hat{a}^\dagger\hat{a}\hat{a}^\dagger:]
    &=\Tr[\hat{\rho}\hat{a}^{\dagger2}\hat{a}^2]
    =\int_{\mathbb{C}^{2}} P(\alpha,\beta)\Tr[\hat{\Lambda}(\alpha,\beta)\hat{a}^{\dagger2}\hat{a}^2]d\alpha d\beta \\
    &=\int_{\mathbb{C}^{2}} P(\alpha,\beta)\frac{\langle \beta^*|\hat{a}^{\dagger2}\hat{a}^2|\alpha\rangle}{\langle\beta^*|\alpha\rangle}d\alpha d\beta
    =\int_{\mathbb{C}^{2}} P(\alpha,\beta)\beta^2 \alpha^2 d\alpha d\beta.
\end{align}
Therefore, we may be able to directly integrate or estimate it by sampling $(\alpha,\beta)$ from $P$ and averaging $\alpha^2\beta^2$ over the samples.
The same procedure works for any normal-ordered operator $f(\hat{a},\hat{a}^\dagger)$ by replacing $\hat{a}\to\alpha$ and $\hat{a}^\dagger\to \beta$.
Especially when the operator is written as a function of $\hat{n}$, $g(\hat{n})$, then the corresponding phase-variable is given by $g(\alpha\beta)$.

When we apply a linear-optical circuit to a given input state, the quantum state transforms as
\begin{align}
    \hat{\rho}\to \hat{U}\hat{\rho}\hat{U}^\dagger,
\end{align}
which is written for the positive $P$-representation as
\begin{align}
    \hat{\rho}=
    \int_{\mathbb{C}^{2M}} P(\bm{\alpha},\bm{\beta})\hat{\Lambda}(\bm{\alpha},\bm{\beta})d\bm{\alpha}d\bm{\beta}~\to~
    \hat{\rho}=
    \int_{\mathbb{C}^{2M}} P(\bm{\alpha},\bm{\beta})\hat{\Lambda}(U\bm{\alpha},U^*\bm{\beta})d\bm{\alpha}d\bm{\beta}.
\end{align}
Therefore, sampling from the positive $P$-representation of the output state of the linear-optical circuit is equivalent to sampling from the positive $P$-representation of the input state and transform the phase variable as $(\bm{\alpha},\bm{\beta})\to (\bm{\alpha}',\bm{\beta}')=(U\bm{\alpha},U^*\bm{\beta})$.

\section{Positive $P$-representation of a squeezed vacuum state}
In this section, we show that a positive $P$-representation of a squeezed vacuum state can be chosen as \cite{janszky1990squeezing}
\begin{align}\label{eq:sq_P}
    P(x, y)
    &=\frac{\sqrt{1+\gamma}}{\pi \gamma}\exp\left(-(x^2+y^2)(\gamma^{-1}+1/2)+xy\right),
\end{align}
where $x$ and $y$ are real numbers.
We follow the proof in Ref.~\cite{janszky1990squeezing}.
We emphasize that this function is defined on the real plane, which is a distinct property of a squeezed state, and that generally the positive $P$-representation is defined on the complex-variable domain:
\begin{align}
    \hat{\rho}=
    \int_{\mathbb{C}^{2M}} P(\bm{\alpha},\bm{\beta})\hat{\Lambda}(\bm{\alpha},\bm{\beta})d\bm{\alpha}d\bm{\beta},
    ~~~
    \hat{\Lambda}(\bm{\alpha},\bm{\beta})\equiv \frac{|\bm{\alpha}\rangle\langle \bm{\beta}^*|}{\langle \bm{\beta}^*|\bm{\alpha}\rangle},
\end{align}
where $|\alpha\rangle$ and $|\beta^*\rangle$ are coherent states.
The claim that we now prove is that the quantum state $|\psi\rangle$ written as
\begin{align}\label{eq:F}
    |\psi\rangle=\int_{\mathbb{R}}dx F(x)|x\rangle ~~~ \text{with} ~~~ F(x)=\frac{(1+\gamma)^{1/4}}{\sqrt{\pi \gamma}}\exp(-x^2/\gamma),
\end{align}
is a single-mode squeezed state.
Using
\begin{align}
    \langle \alpha|\beta\rangle=\exp\left(\alpha^*\beta-\frac{1}{2}|\alpha|^2-\frac{1}{2}|\beta|^2\right),
\end{align}
Eq.~\eqref{eq:F} leads to Eq.~\eqref{eq:sq_P}. 
Thus, it suffices to show the claim to prove that Eq.~\eqref{eq:sq_P} is a positive $P$-representation of a squeezed vacuum state.

To show that $|\psi\rangle$ is a squeezed vacuum state, we consider the characteristic function of $|\psi\rangle$,
\begin{align}
    \chi_\psi(\eta)=\text{Tr}\left[|\psi\rangle\langle\psi|\exp(\eta \hat{a}^\dagger-\eta^*\hat{b})\right]
    &=\int_{\mathbb{R}^2}dx dy F(x)F(y)\exp\left[-\frac{1}{2}|\eta|^2+\eta y-\eta^*x-\frac{1}{2}(x-y)^2\right] \\ 
    &=\frac{\sqrt{1+\gamma}}{\pi\gamma}\int_{\mathbb{R}^2}dx dy \exp\left[-\frac{x^2+y^2}{\gamma}-\frac{1}{2}|\eta|^2+\eta y-\eta^*x-\frac{1}{2}(x-y)^2\right] \\ &=\exp[-\frac{|\eta \gamma-\eta^*(\gamma+2)|^2}{8(1+\gamma)}].
\end{align}
By changing the variable with real numbers, $\eta\equiv q+ip$, we obtain the following expression:
\begin{align}
    \chi_\psi(q,p)=\exp[-\frac{q^2+(1+\gamma)^2p^2}{2(1+\gamma)}].
\end{align}
Thus, it is the characteristic function of a multivariate normal distribution of a covariance matrix,
\begin{align}
    \Sigma=\text{diag}\left(\frac{1}{1+\gamma},1+\gamma\right).
\end{align}
Since the covariance matrix of a $p$-squeezed vacuum state with a squeezing parameter $r$ is $\text{diag}(e^{2r},e^{-2r})$, it shows that $|\psi\rangle$ is a $p$-squeezed vacuum state and the relation $\gamma+1=e^{2r}$.

We note that the above expression holds only if $r>0$.
Especially when $r=0$, i.e., when the state is vacuum, the expression is not well defined, but the positive $P$-representation can be obtained by using Eq.~\eqref{eq:q}:
\begin{align}\label{eq:vacuum}
    P_0(\alpha,\beta)&=\frac{1}{4\pi^2}\exp[-\frac{1}{2}(|\alpha|^2+|\beta|^2)],
\end{align}
where $\alpha$ and $\beta$ are complex.
Since the $P$-representation is still Gaussian, one may easily generalize the analytical result of the Fourier components $\tilde{G}(k)$ for the vacuum, while nonzero-squeezing-parameter cases are more generic for molecular vibronic spectra.
In addition, for the purpose of the Gaussian boson sampling setting, or the molecular vibronic spectra, squeezing parameters $r>0$ is sufficient because the rotation part can be moved to the linear-optical circuit.
Nonetheless, squeezed vacuum states in other directions can be obtained by simply rotating the state) as
\begin{align}
    \hat{\rho}\to \hat{R}(\phi)\hat{\rho}\hat{R}^\dagger(\phi)=\int_{\mathbb{R}^2}P(\alpha,\beta)\hat{\Lambda}(\alpha e^{i\phi},\beta e^{-i\phi})d\alpha d\beta,
\end{align}
where $\hat{R}(\phi)=e^{-i\hat{a}^\dagger\hat{a}\phi}$ is the rotation operator.
Thus, if we change the variables, the domain for the integral may not be the real plane.

\section{Exact solution of Fourier components of molecular vibronic spectra}
Here, we show the derivation of the analytic solution of Fourier components of the vibronic spectra.
Let us first write the initial positive $P$-representation as (see Sec.~\ref{sec:P} for more details.) \cite{janszky1990squeezing}
\begin{align}
    P_\text{in}(\bm{x},\bm{y})
    &=\prod_{i=1}^M\left[\frac{\sqrt{1+\gamma_i}}{\pi \gamma_i}\exp\left(-(x_i^2+y_i^2)(\gamma_i^{-1}+1/2)+x_iy_i\right)\right] \\ 
    &=\mathcal{N}\exp\left[-\bm{x}. A. \bm{x}-\bm{y}. B.\bm{y}-\bm{x}.C.\bm{y}-\bm{y}.C^\text{T}.\bm{x}\right],
\end{align}
where we define
\begin{align}
    A\equiv\text{diag}(\gamma_i^{-1}+1/2)_{i=1}^M, ~~~B\equiv\text{diag}(\gamma_i^{-1}+1/2)_{i=1}^M, 
    ~~~C\equiv-\mathbb{1}/2,~~~
    \mathcal{N}\equiv\prod_{i=1}^M\frac{\sqrt{1+\gamma_i}}{\pi \gamma_i}.
\end{align}
Here, $\bm{\alpha}$ and $\bm{\beta}$ are real $M$-dimensional vectors.
From the main text, we have found the expression of the Fourier components of molecular vibronic spectra,
\begin{align}
    \tilde{G}(k)=\langle e^{-ik\theta \hat{\bm{n}}\cdot \bm{\omega}}\rangle,
\end{align}
where the expectation value is over the output state of Gaussian boson sampling.
By using the relation $e^{-\gamma \hat{n}}=:e^{\hat{n}(e^{-\gamma}-1)}:$ \cite{vargas2006normal}, we can write the operators as a normal-ordered form:
\begin{align}
    \tilde{G}(k)=\left\langle :\exp\left[\sum_{i=1}^M \hat{n}_i(e^{-ik\theta \omega_i}-1)\right]:\right\rangle.
\end{align}

Now, we compute the expectation value,
\begin{align}
    \tilde{G}(k)=\left\langle :\exp\left[\sum_{i=1}^M \hat{n}_i(e^{i\phi_i}-1)\right]:\right\rangle
    &=\left\langle \exp\left[\sum_{i=1}^M x_i'y_i'(e^{i\phi_i}-1)\right]\right\rangle_P \\ 
    &=\left\langle \exp\left[\sum_{i=1}^M \left(\sum_{j=1}^MU_{ij}x_j+x_{0i}\right)\left(\sum_{k=1}^MU_{ik}^*y_k+y_{0i}\right)(e^{i\phi_i}-1)\right]\right\rangle_P \\
    &=\left\langle \exp\left[\bm{x}.\mathcal{U}.\bm{y}+\bm{a}.\bm{x}+\bm{b}.\bm{y}+c_0\right]\right\rangle_P \\ 
    &=\mathcal{N}\int d\bm{x} d\bm{y}\exp\left[-\bm{x}. A. \bm{x}-\bm{y}. B.\bm{y}-\bm{x}.\tilde{C}.\bm{y}-\bm{y}.\tilde{C}^\text{T}.\bm{x}+\bm{a}.\bm{x}+\bm{b}.\bm{y}+c_0\right] \\ 
    &=\mathcal{N}\int d\bm{q}\exp\left[-\bm{q}.Q.\bm{q}/2+\bm{c}.\bm{q}+c_0\right] \label{eq:Q_mat},
\end{align}
where $\phi_i\equiv -k\theta \omega_i$,
\begin{align}
    &\mathcal{U}\equiv
    U^\text{T}\Phi U^*,~~~
    \Phi\equiv
    \text{diag}(e^{i\phi_1}-1,\dots,e^{i\phi_M}-1),
    ~~~
    \tilde{C}\equiv C-\mathcal{U}/2,~~~
    Q=2
    \begin{pmatrix}
        A & \tilde{C} \\ 
        \tilde{C}^\text{T} & B
    \end{pmatrix}, \\
    &\bm{c}\equiv
    \begin{pmatrix}
        \bm{a} \\
        \bm{b}
    \end{pmatrix},~~~
    \bm{a}=U^\text{T}\Phi \bm{y}_0,~~~
    \bm{b}=U^\dagger\Phi \bm{x}_0,~~~
    c_0=\bm{x}_0\Phi \bm{y}_0,
\end{align}
and $\langle \cdot \rangle_P$ is the average over the positive $P$-representation.
We note that the displacement occurs after the linear-optical unitary, which matches the displacement $\bm{\delta}/\sqrt{2}$ in the Doktorov transformation, i.e., $\bm{x}_0=\bm{y}_0^*=\bm{\delta}/\sqrt{2}$
Here, we note that $Q$'s real part is positive-definite.
To see this, we decompose $Q$ as
\begin{align}
    Q=
    2\mathbb{1}_2\otimes \Gamma^{-1}
    +
    \begin{pmatrix}
        \mathbb{1}_M & -\mathbb{1}_M \\ 
        -\mathbb{1}_M & \mathbb{1}_M
    \end{pmatrix}
    -
    \begin{pmatrix}
        0 & \mathcal{U} \\ 
        \mathcal{U}^\text{T} & 0
    \end{pmatrix}
    =
    2\mathbb{1}_2\otimes \Gamma^{-1}
    +
    \begin{pmatrix}
        \mathbb{1}_M & -U^\text{T}\text{diag}(e^{i\phi_j})_{j=1}^MU^* \\ 
        -U^\dagger\text{diag}(e^{i\phi_j})_{j=1}^MU & \mathbb{1}_M
    \end{pmatrix},
\end{align}
where $\Gamma\equiv \text{diag}(\gamma_1,\dots,\gamma_M)$.
Now, we can write the real part of $Q$ as
\begin{align}
    \text{Re}(Q)=2\mathbb{1}_2\otimes \Gamma^{-1}+
    \begin{pmatrix}
        \mathbb{1}_M & L \\ 
        L^\text{T} & \mathbb{1}_M
    \end{pmatrix},
\end{align}
where $L\equiv \text{Re}(-U^\text{T}\text{diag}(e^{i\phi_j})_{j=1}^MU^*)$.
Since $\|L\|_\text{op}\leq 1$, $\text{Re}(Q)$ is positive-definite \cite{bhatia2009positive}.
Also, it is well known that the Gaussian integral in Eq.~\eqref{eq:Q_mat} converges if the real part of the complex symmetric matrix $Q$ is positive-definite.
Therefore, the integral always converges.

Using the Gaussian integral formula,
\begin{align}
    \int d\bm{q}\exp[-\bm{q}.Q.\bm{q}/2+\bm{c}.\bm{q}]
    =\sqrt{\frac{(2\pi)^{2M}}{|Q|}}\exp\left(\frac{1}{2}\bm{c}.Q^{-1}.\bm{c}\right),
\end{align}
we obtain the analytical form of the Fourier components,
\begin{align}\label{eq:analytic}
    \tilde{G}(k)=\mathcal{N}\sqrt{\frac{(2\pi)^{2M}}{|Q|}}\exp\left(\frac{1}{2}\bm{c}.Q^{-1}.\bm{c}+c_0\right),
\end{align}
where
\begin{align}
    Q=
    \begin{pmatrix}
        2\Gamma^{-1}+\mathbb{1}_M & -U^\text{T}\text{diag}(e^{i\phi_j})_{j=1}^MU^* \\ 
        -U^\dagger\text{diag}(e^{i\phi_j})_{j=1}^MU & 2\Gamma^{-1}+\mathbb{1}_M
    \end{pmatrix}.
\end{align}
We note that for simplicity, we have assumed nonzero squeezing, which is generic for molecular vibronic spectra, while a similar expression can be obtained when there is a zero squeezing because the positive $P$-function is still Gaussian (See Eq.~\eqref{eq:vacuum}).

\begin{figure}[t]
\includegraphics[width=300px]{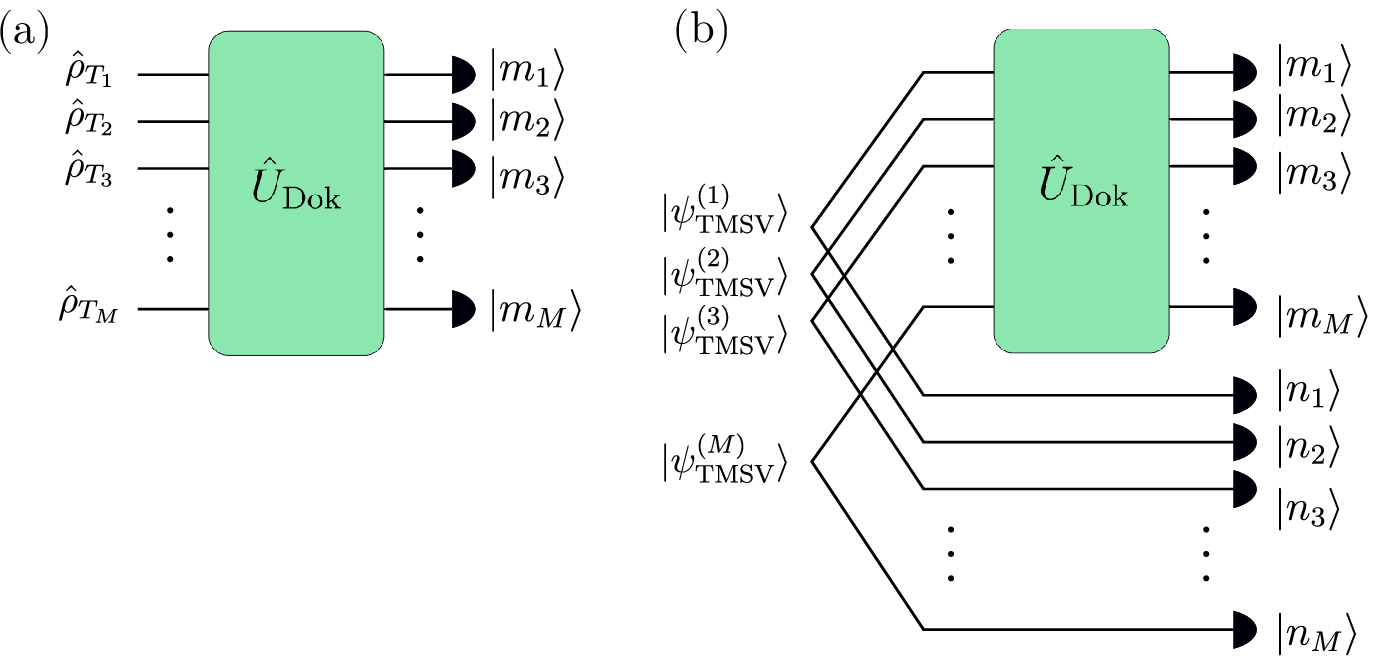}
\caption{Transformation of molecular vibronic spectra (a) at finite temperature to molecular vibronic spectra at zero temperature.
(b) By introducing additional $M$ modes and two-mode squeezing operations, one may treat the problem as at zero temperature.
For finite temperature case, the cost is only the additional $M$ modes and we can exploit the same solution as at zero temperature.
}
\label{fig:finite_T3}
\end{figure}

\section{Exact solution for molecular vibronic spectra at finite temperature}
We now show that our method is applicable for molecular vibronic spectra even at finite temperature.
At finite temperature $T$, the molecular vibronic spectra is written as
\begin{align}\label{eq:FCP_T}
    \text{FCP}(\Omega)
    &=\sum_{\bm{n},\bm{m}=\bm{0}}^\infty p(\bm{n},\bm{m})\delta\left(\Omega-(\bm{\omega}'\cdot\bm{m}-\bm{\omega}\cdot\bm{n})\right),
\end{align}
where
\begin{align}
    p(\bm{n},\bm{m})=p_T(\bm{n})|\langle \bm{n}|\hat{U}_\text{Dok}|\bm{m}\rangle|^2.
\end{align}
Here, $p_T(\bm{n})\equiv \langle \bm{n}|\hat{\rho}_T|\bm{n}\rangle$ is the probability of photon number $|\bm{n}\rangle$ from an $M$-mode thermal state $\hat{\rho}_T$ at temperature $T$ (in general we do not need to assume the same temperature.).
In Ref.~\cite{huh2017vibronic}, it was shown that one may construct a Gaussian boson sampling for the finite temperature case.
The key idea is to add $M$ auxiliary modes and prepare two-mode squeezed states to simulate thermal states and detect the additional modes as well, which is depicted in Fig.~\ref{fig:finite_T3}.
In this case, the original $M$ modes' output photons correspond to $\bm{m}$ and the additional $M$ modes' output photons corresponds to $\bm{n}$.
Therefore, we can again use the same framework to obtain the solution.
We can rewrite the FCP as
\begin{align}\label{eq:FCP_T2}
    \text{FCP}(\Omega)=\sum_{\bm{m},\bm{n}=\bm{0}}^\infty |\langle \bm{m},\bm{n}|\hat{U}_\text{Dok}'|\bm{0}\rangle|^2\delta\left(\Omega-(\bm{\omega}'\cdot\bm{m}-\bm{\omega}\cdot\bm{n})\right),
\end{align}
which is the same form as the zero temperature case except for the negative part in delta function.
Here, $\hat{U}'_\text{Dok}=\hat{U}_\text{Dok}\hat{U}_\text{TMSV}$ is obtained by applying two-mode squeezing operators.
Again, the Fourier transform of the FCP at finite temperature is given by
\begin{align}
    \tilde{G}(k)=\langle e^{-ik\hat{n}\cdot\bm{\omega}''\theta}\rangle,
\end{align}
where $\bm{\omega}''=({\bm{\omega}'}^{\text{T}},-\bm{\omega}^\text{T})^\text{T}$.
Therefore, the analytic solution for Fourier components can easily be found.
Note that without introducing additional modes, one can directly use the positive $P$-representation for a squeezed thermal input state and follow the same procedure as the zero temperature case.

\section{Computation of low-rank loop hafnian}
Here, we generalize the algorithm of computing the hafnian of a low-rank symmetric matrix to loop hafnian \cite{bjorklund2019faster}.
The relevant matrix of which we compute the loop hafnian is $\tilde{\Sigma}$, which is obtained by replacing the diagonal elements of a symmetric $\Sigma$ by $\mu$ vector, and we assume that the symmetric matrix $\Sigma$ has a rank $r$; thus, we can find $G\in \mathbb{C}^{n\times r}$ such that $GG^\text{T}=\Sigma$.
We then introduce indeterminates $x_1,\dots,x_r$, and consider the multivariate polynomial
\begin{align}
    q(x_1,\dots,x_r)=\prod_{i=1}^n\left(\sum_{j=1}^rg_{i,j}x_j+\mu_j\right).
\end{align}
Note that the difference from the hafnian case is the additional term $\mu_j$.
We let $\mathcal{P}_{2s,r}$ be the set of integer $r$-partitions of $2s$, i.e., tuples $(p_1,p_2,\dots,p_r)$ such that $p_i\geq0$ for all $i$ and $\sum_i p_i=2s$, and let $\mathcal{E}_{2s,r}$ be the subset of $\mathcal{P}_{s,r}$ such that $p_i$ is even for all $i$.
We then expand the polynomial and identify the coefficients $\lambda_p$ as
\begin{align}
    \tilde{q}(x_1,\dots,x_r)=\sum_{s=0}^{[n/2]}\sum_{p=(p_1,\dots,p_{2s})\in \mathcal{P}_{2s,r}}\lambda_p\prod_{i=1}^r x_i^{p_i}.
\end{align}
where we dropped a polynomial composed of odd number of $x_i$'s from $q(x_1,\dots,x_r)$.
Since
\begin{align}
    |\mathcal{P}_{2s,r}|=\binom{2s+r-1}{r-1} ~~~ \text{and} ~~~ |\mathcal{E}_{2s,r}|=\binom{s+r-1}{r-1},
\end{align}
The coefficients $\lambda_p$ can be found in $|\mathcal{P}_{n,r}|\text{poly}(n)$ running time.
Meanwhile, the loop hafnian can be written as
\begin{align}
    \lhaf(\tilde{\Sigma})=\sum_{s=0}^{[n/2]}\sum_{e\in\mathcal{E}_{2s,r}}\lambda_e\prod_{i=1}^r(e_i-1)!!,
\end{align}
where $a!!=1$ if $a\leq 0$.
For given coefficients $\lambda_e$, the loop hafnian can be computed in $|\mathcal{E}_{n,r}|\poly(n)$ time.
Thus, when $\Sigma$ is a rank $r$ matrix, one can compute its relevant loop hafnian $\lhaf(\tilde{\Sigma})$ can be evaluated in $\binom{n+r-1}{r-1}\poly(n)$.

\section{Fourier components of molecular vibronic spectra}\label{sec:hafnian}
In this section, we simplify the Fourier component of molecular vibronic spectra.
We focus on the displaced squeezed Fock state input $\hat{D}(\bm{\alpha})\hat{S}(\bm{r}_0)|\bm{n}\rangle$.
In this case, the Fourier component can be written as
\begin{align}
    \tilde{G}(k)
    &=\langle\bm{n}|\hat{S}^\dagger(\bm{r}_0)\hat{D}^\dagger(\bm{\alpha})\hat{U}^\dagger e^{-ik\theta \hat{\bm{n}}\cdot\bm{\omega}}\hat{U}\hat{D}(\bm{\alpha})\hat{S}(\bm{r}_0)|\bm{n}\rangle
    \equiv \langle\bm{n}|\hat{W}|\bm{n}\rangle
    =\langle \bm{n}|\hat{D}(\bm{\xi})\hat{U}_\text{lin2}\hat{S}(\bm{r})\hat{U}_\text{lin1}|\bm{n}\rangle,
\end{align}
where we have used Bloch-Messiah decomposition of the Gaussian unitary matrix $\hat{W}$.

Before we derive the most general case's formula, we first consider a simpler case.
First, we consider $\langle n|\hat{S}(r)|n\rangle$ for a single mode:
\begin{align}\label{eq:nsn}
    \langle n|\hat{S}(r)|n\rangle
    =\frac{1}{\pi^2}\int d\alpha d\beta \langle n|\alpha\rangle\langle \alpha|\hat{S}(r)|\beta\rangle\langle \beta|n\rangle
    =\frac{1}{n!\pi^2}\int d\alpha d\beta \alpha^n\beta^{*n}\exp\left[-\frac{1}{2}\left(|\alpha|^2+|\beta|^2\right)\right]\langle \alpha|\hat{S}(r)|\beta\rangle.
\end{align}
Note that
\begin{align}
    \langle \alpha|\hat{S}(r)|\beta\rangle
    &=\frac{1}{\sqrt{\cosh{r}}}\exp\bigg[-\frac{1}{2}\bigg(|\alpha-\beta\cosh{r}-\beta^*\sinh{r}|^2-\tanh{r}(\alpha^*-\beta^*\cosh{r}-\beta \sinh{r})^2 \nonumber \\ 
    &~~~+\alpha (\beta^* \cosh{r}+\beta\sinh{r})-\alpha^*(\beta \cosh{r}+\beta^*\sinh{r})\bigg)\bigg] \\
    &=\frac{1}{\sqrt{\cosh{r}}}\exp\bigg[-\frac{1}{2}
    v(\alpha,\beta)^\text{T}.
    \begin{pmatrix}
        0 & 1/2 & 0 & 0 \\
        1/2 & -\tanh{r} & -\sech{r} & 0 \\ 
        0 & -\sech{r} & \tanh{r} & 1/2\\ 
        0 & 0 & 1/2 & 0
    \end{pmatrix}.
    v(\alpha,\beta)\bigg].
\end{align}
Here, we defined $v(\alpha,\beta)\equiv (\alpha,\alpha^*,\beta,\beta^*)^\text{T}$.
Therefore, Eq.~\eqref{eq:nsn} can be rewritten as
\begin{align}
    \langle n|\hat{S}(r)|n\rangle=\frac{1}{n!\pi^2\sqrt{\cosh{r}}}\int d\alpha d\beta \alpha^n\beta^{*n}\exp\left[
    -\frac{1}{2}
    v(\alpha,\beta)^\text{T}.
    \Sigma_s^{-1}.
    v(\alpha,\beta)
    \right],
\end{align}
where we have defined
\begin{align}
    \Sigma_s^{-1}\equiv
    \begin{pmatrix}
        0 & 1 & 0 & 0 \\
        1 & -\tanh{r} & -\sech{r} & 0 \\ 
        0 & -\sech{r} & \tanh{r} & 1\\ 
        0 & 0 & 1 & 0
    \end{pmatrix},~~~
    \Sigma_s=
    \begin{pmatrix}
        \tanh{r} & 0 & 1 & \sech{r} \\
        0 & 0 & 0 & 1 \\ 
        1 & 0 & 0 & 0\\ 
        \sech{r} & 1 & 0 & -\tanh{r}
    \end{pmatrix}.
\end{align}
Finally, using Wick's theorem \cite{kan2008moments}, we obtain
\begin{align}
    \langle n|\hat{S}(r)|n\rangle
    =\frac{\langle\alpha^n\beta^{*n}\rangle}{n!\sqrt{\cosh{r}}}
    =\frac{\haf((\Sigma_s)_{n})}{n!\sqrt{\cosh{r}}},
\end{align}
where $(\Sigma_s)_{n}$ is obtained by repeating the first and fourth rows and columns $n$ times, which accounts for $\alpha$ and $\beta^*$, respectively.

Using the same method, we derive a more general case:
\begin{align}
    &\langle \bm{n}|\hat{D}(\bm{\xi})\hat{U}_\text{lin2}\hat{S}(\bm{r})\hat{U}_\text{lin1}|\bm{n}\rangle \\
    &=\frac{1}{\pi^{4M}}\int d\bm{\alpha}d\bm{\gamma}d\bm{\omega}d\bm{\beta}\langle\bm{n}|\hat{D}(\bm{\xi})|\bm{\alpha}\rangle\langle\bm{\alpha}|\hat{U}_\text{lin2}|\bm{\gamma}\rangle\langle{\bm{\gamma}}|\hat{S}(\bm{r})|\bm{\omega}\rangle\langle\bm{\omega}|\hat{U}_\text{lin1}|\bm{\beta}\rangle\langle\bm{\beta}|\bm{n}\rangle\\
    &=\frac{1}{\bm{n}!\pi^{4M}}\int d\bm{\alpha}d\bm{\gamma}d\bm{\omega}d\bm{\beta}\prod_{i=1}^M[(\alpha_i+\xi_i)\beta_i^*] \nonumber \\ 
    &~~~~\times\exp\left[-\frac{1}{2}\left(|\bm{\alpha}+\bm{\xi}|^2+|\bm{\alpha}|^2+2|\bm{\beta}|^2+|\bm{\gamma}|^2+|\bm{\omega}|^2\right)+\bm{\alpha}^*.U_\text{lin2}.\bm{\gamma}+\bm{\omega}^*.U_\text{lin1}.\bm{\beta}+\frac{1}{2}(\bm{\xi}.\bm{\alpha}^*-\bm{\xi}^*.\bm{\alpha})\right]\langle{\gamma}|\hat{S}(\bm{r})|\bm{\omega}\rangle \\ 
    &=\frac{1}{\bm{n}!\pi^{4M}\prod_{i=1}^M\sqrt{\cosh{r_i}}}\int d\bm{\alpha}d\bm{\gamma}d\bm{\omega}d\bm{\beta}\prod_{i=1}^M[(\alpha_i+\xi_i)\beta_i^*]\exp\left[-|\bm{\alpha}|^2-|\bm{\beta}|^2-\frac{1}{2}|\bm{\xi}|^2-\bm{\alpha}.\bm{\xi}^*+\bm{\alpha}^*.U_\text{lin2}.\bm{\gamma}+\bm{\omega}^*.U_\text{lin1}.\bm{\beta}\right] \nonumber \\ 
    &~~~~\times \exp\left[-\frac{1}{2}v(\bm{\gamma},\bm{\omega})^\text{T}.\Sigma_s^{-1}.v(\bm{\gamma},\bm{\omega})\right] \\ 
    &=\frac{1}{\bm{n}!\pi^{4M}}\int d\bm{\alpha}d\bm{\gamma}d\bm{\omega}d\bm{\beta}\prod_{i=1}^M[(\alpha_i+\xi_i)\beta_i^*]\exp\left(-\frac{1}{2}|\bm{\xi}|^2-\bm{\alpha}.\bm{\xi}^*\right)\exp\left(-\frac{1}{2}v(\bm{\alpha},\bm{\beta},\bm{\gamma},\bm{\omega})^\text{T}.\Sigma^{-1}.v(\bm{\alpha},\bm{\beta},\bm{\gamma},\bm{\omega})\right) \\ 
    &=\frac{1}{\bm{n}!\pi^{4M}Z}\int d\bm{\alpha}d\bm{\gamma}d\bm{\omega}d\bm{\beta}\prod_{i=1}^M[(\alpha_i+\xi_i+\zeta_i^{\alpha})(\beta_i+\zeta_i^{\beta^*})^*]^{n_i}\exp\left(-\frac{1}{2}v(\bm{\alpha},\bm{\beta},\bm{\gamma},\bm{\omega})^\text{T}.\Sigma^{-1}.v(\bm{\alpha},\bm{\beta},\bm{\gamma},\bm{\omega})\right) \\ 
    &=\frac{\lhaf(\tilde{\Sigma}_{\bm{n}})}{\bm{n}!Z}.
\end{align}
Here, we have introduced various parameters, which are defined as follows:
First, we have defined a complex vector as
\begin{align}
    v(\bm{\alpha},\bm{\beta},\bm{\gamma},\bm{\omega})
    =(\alpha_1,\dots,\alpha_M,\alpha_1^*,\dots,\alpha_M^*,\beta_1,\dots,\beta_M,\beta_1^*,\dots,\beta_M^*,\gamma_1,\dots,\gamma_M,\gamma_1^*,\dots,\gamma_M^*,\omega_1,\dots,\omega_M,\omega_1^*,\dots,\omega_M^*),
\end{align}
and the complex symmetric covariance matrix $\Sigma$ as 
\begin{align}
    \Sigma^{-1}=
    \begin{pmatrix}
        \Sigma_x & -\mathcal{W} \\ 
        -\mathcal{W}^\text{T} & \Sigma_s^{-1}
    \end{pmatrix},~~~
    \mathcal{U}_\text{lin1}=
    \begin{pmatrix}
        0 & U_\text{lin1}^\text{T} \\
        0 & 0
    \end{pmatrix},~~~
    \mathcal{U}_\text{lin2}=
    \begin{pmatrix}
        0 & 0 \\
        U_\text{lin2} & 0
    \end{pmatrix},~~~
    \mathcal{W}=
    \begin{pmatrix}
        \mathcal{U}_\text{lin2} & 0 \\ 
        0 & \mathcal{U}_\text{lin1}    
    \end{pmatrix},
    \\
    \Sigma_x=
    \begin{pmatrix}
        \sigma_x \otimes \mathbb{1} & 0 \\
        0 & \sigma_x \otimes \mathbb{1}
    \end{pmatrix},~~~
        \Sigma_s=
    \begin{pmatrix}
        \tanh{\bm{r}} & \mathbb{1}_M & 0 & \sech{\bm{r}} \\ 
        \mathbb{1}_M & 0 & 0 & 0 \\ 
        0 & 0 & 0 & \mathbb{1}_M \\ 
        \sech{\bm{r}} & 0 & \mathbb{1}_M & -\tanh{\bm{r}}
    \end{pmatrix},
\end{align}
where $\sigma_x$ is the Pauli $x$ matrix.
Here, using the equality
\begin{align}
    \begin{pmatrix}
        A & B \\
        C & D
    \end{pmatrix}^{-1}
    =
    \begin{pmatrix}
    A^{-1}+A^{-1}B(D-CA^{-1}B)^{-1}CA^{-1} & -A^{-1}B(D-CA^{-1}B)^{-1} \\ 
    -(D-CA^{-1}B)^{-1}CA^{-1} & (D-CA^{-1}B)^{-1}
    \end{pmatrix},
\end{align}
we obtain
\begin{align}
    \Sigma=
    \begin{pmatrix}
        U_\text{lin2}\tanh{\bm{r}}U_\text{lin2}^\text{T} & \mathbb{1} & 0 & U_\text{lin2}\sech{\bm{r}}U_\text{lin1} & U_\text{lin2}\tanh{\bm{r}} & U_\text{lin2} & 0 & U_\text{lin2}\sech{\bm{r}} 
        \\ 
        \mathbb{1} & 0 & 0 & 0 & 0 & 0 & 0 & 0 
        \\ 
        0 & 0 & 0 & \mathbb{1} & 0 & 0 & 0 & 0 
        \\ 
        U_\text{lin1}^\text{T}\sech{\bm{r}}U_\text{lin2}^\text{T} & 0 & \mathbb{1} & -U_\text{lin1}^\text{T}\tanh{\bm{r}}U_{\text{lin1}} & U_\text{lin1}^\text{T}\sech{\bm{r}} & 0 & U_\text{lin1}^\text{T} & -U_\text{lin1}^\text{T}\tanh{\bm{r}}
        \\ 
        \tanh{\bm{r}} U_\text{lin2}^\text{T} & 0 & 0 & \sech{\bm{r}}U_\text{lin1} & \tanh{\bm{r}} & \mathbb{1} & 0 & \sech{\bm{r}} \\ 
        U_\text{lin2}^\text{T} & 0 & 0 & 0 & \mathbb{1} & 0 & 0 & 0 \\ 
        0 & 0 & 0 & U_\text{lin1} & 0 & 0 & 0 & \mathbb{1} \\ 
        \sech{\bm{r}}U_{\text{lin2}}^\text{T} & 0 & 0 & -\tanh{\bm{r}}U_{\text{lin1}} & \sech{\bm{r}} & 0 & \mathbb{1} & -\tanh{\bm{r}}
    \end{pmatrix}.
\end{align}
Note that the relevant part for the loop hafnian is the ones that are related to $\bm{\alpha}$ and $\bm{\beta}^*$ (for this reason, we dropped other parts in the main text), which is
\begin{align}
    \Sigma_{\bm{\alpha},\bm{\beta}^*}=
    \begin{pmatrix}
        U_\text{lin2}\tanh{\bm{r}}U_\text{lin2}^\text{T} & U_\text{lin2}\sech{\bm{r}}U_\text{lin1} 
        \\ 
        U_\text{lin1}^\text{T}\sech{\bm{r}}U_\text{lin2}^\text{T} & -U_\text{lin1}^\text{T}\tanh{\bm{r}}U_{\text{lin1}}
    \end{pmatrix},
\end{align}
and $\tilde{\Sigma}_{\bm{n}}$ is the matrix that is obtained by replacing the diagonal element of $\Sigma_{\bm{\alpha},\bm{\beta}^*}$ by $(\bm{\xi}+\bm{\zeta}^{\alpha},\bm{\zeta}^{\beta^*})$ and repeating $i$th row and column of each block matrix by $n_i$.
Here,
\begin{align}
    \bm{\zeta}
    =-\Sigma.\bm{\xi}_\text{ex}^*,~~~
    \begin{pmatrix}
        \bm{\zeta}^\alpha \\ \bm{\zeta}^{\beta^*}
    \end{pmatrix}
    =-
    \begin{pmatrix}
        U_\text{lin2}\tanh{\bm{r}}U_\text{lin2}^\text{T}.\bm{\xi}^* \\ U_\text{lin1}^\text{T}\sech{\bm{r}}U_\text{lin2}^\text{T}.\bm{\xi}^*
    \end{pmatrix},
\end{align}
where $\bm{\xi}_\text{ex}^*$ is interpreted as an extended vector for other parts than $\bm{\alpha}$ such as $\bm{\alpha}^*$, $\bm{\beta}$ to be zero, i.e., $\bm{\xi}^*_\text{ex}=(\bm{\xi}^*,0,\dots,0)$, and
$(\bm{\zeta}^{\alpha}, \bm{\zeta}^{\beta^*})$ are restriction to the space of $\bm{\alpha}$ and $\bm{\beta}^*$.
Finally, the normalization factor $Z$ can be compactly written as
\begin{align}
    Z^{-1}=\langle 0|\hat{D}(\bm{\xi})\hat{U}_\text{lin2}\hat{S}(\bm{r})\hat{U}_\text{lin1}|0\rangle
    =\exp\left(\frac{1}{2}\bm{\xi}_\text{ex}^\text{T}\Sigma\bm{\xi}_\text{ex}\right)/\prod_{i=1}^M \sqrt{\cosh{r_i}},
\end{align}
which gives the analytic expression for the zero temperature case as we obtained in Eq.~\eqref{eq:analytic}.

We note that
\begin{align}\label{eq:norm}
    \Sigma_{\bm{\alpha},\bm{\beta}^*}=
    \begin{pmatrix}
        U_{\text{lin2}} & 0 \\ 
        0 & U_\text{lin1}^\text{T}
    \end{pmatrix}
    \begin{pmatrix}
        \tanh{\bm{r}} & \sech{\bm{r}} \\ 
        \sech{\bm{r}} & -\tanh{\bm{r}}
    \end{pmatrix}
    \begin{pmatrix}
        U_{\text{lin2}}^\text{T} & 0 \\ 
        0 & U_\text{lin1}
    \end{pmatrix}
\end{align}
and that the operator norm is $1$ regardless of $U_\text{lin1}$, $U_\text{lin2}$ and $\bm{r}$.
A more general matrix element such as $\langle \bm{n}|\hat{D}(\bm{\xi})\hat{U}_\text{lin2}\hat{S}(\bm{r})\hat{U}_\text{lin1}|\bm{m}\rangle$ can be similarly obtained \cite{quesada2019franck}.

As a remark, when $\bm{r}=0$, the relevant matrix becomes
\begin{align}
    \begin{pmatrix}
        0 & U_{\text{lin2}}U_{\text{lin1}} \\ 
        U_{\text{lin2}}^\text{T}U_{\text{lin1}}^\text{T} & 0
    \end{pmatrix}.
\end{align}
Since for a matrix $A$,
\begin{align}
    \haf
    \begin{pmatrix}
        0 & A \\ 
        A^T & 0
    \end{pmatrix}
    =\per A
\end{align}
we can see that when squeezing parameters are zero, we recover the Fock-state boson sampling case.

\section{Approximation algorithm of hafnian}
We provide an algorithm to approximate a hafnian.
First, from Lemma 1 in Ref.~\cite{kan2008moments},
\begin{align}
    x_1^{n_1}x_2^{n_2}\cdots x_M^{n_M}=\frac{1}{n!}\sum_{v_1=0}^{n_1}\cdots \sum_{v_M=0}^{n_M}(-1)^{\sum_{i=1}^M v_i}\binom{n_1}{v_1}\cdots \binom{n_M}{v_M}
    \left(\sum_{i=1}^M h_i x_i\right)^n,
\end{align}
where $h_i=n_i/2-v_i$.
Using the lemma, Proposition 1 in Ref.~\cite{kan2008moments} states that for $z=(z_1,\dots,z_M)^\text{T}\sim \mathcal{N}(0,\Sigma)$, where $\Sigma$ is a real symmetric matrix and $z$ can be a complex variable vector,
\begin{align}\label{eq:haf_sum}
    \haf(\Sigma_{\bm{n}})=\mathbb{E}_z\left[\prod_{i=1}^M z_i^{n_i}\right]=\frac{1}{(n/2)!}\sum_{v_1=0}^{n_1}\cdots \sum_{v_M=0}^{n_M}(-1)^{\sum_{i=1}^M v_i}\binom{n_1}{v_1}\cdots\binom{n_M}{v_M}\left(\frac{h^\text{T}\Sigma h}{2}\right)^{n/2},
\end{align}
where $n=n_1+\cdots+n_M$ and $h=(n_1/2-v_1,\dots,n_M/2-v_M)^\text{T}$ and $n$ is even.
We note that the equality between the hafnian and the sum in Eq.~\eqref{eq:haf_sum} holds even for general complex symmetric matrices.
Especially when $n_i=1$ for all $i\in \{1,\dots,n\}$,
\begin{align}
    \mathbb{E}_z\left[\prod_{i=1}^{n} z_i\right]
    &=\frac{1}{(n/2)!}\sum_{v_1=0}^{1}\cdots \sum_{v_{M}=0}^{1}(-1)^{\sum_{i=1}^{n} v_i}\left(\frac{h^\text{T}\Sigma h}{2}\right)^{n/2}
    =\frac{1}{2^{n}}\sum_{v_1=0}^{1}\cdots \sum_{v_{n}=0}^{1}(-1)^{\sum_{i=1}^{n} v_i}\frac{2^{n}}{(n/2)!}\left(\frac{h^\text{T}\Sigma h}{2}\right)^{n/2} \\ 
    &=\mathbb{E}_{v}\left[(-1)^{\sum_{i=1}^{n} v_i}\frac{2^{n/2}}{(n/2)!}\left(h^\text{T}\Sigma h\right)^{n/2}\right],
\end{align}
where $h=(1/2-v_1,\dots,1/2-v_M)^\text{T}$.
In this case, $M=n$.
Therefore, 
\begin{align}
    (-1)^{\sum_{i=1}^{n} v_i}\frac{2^{n/2}}{(n/2)!}\left(h^\text{T}\Sigma h\right)^{n/2}
\end{align}
is an unbiased estimator for the hafnian in the sense that
\begin{align}
    \mathbb{E}_v\left[(-1)^{\sum_{i=1}^{n} v_i}\frac{2^{n/2}}{(n/2)!}\left(h^\text{T}\Sigma h\right)^{n/2}\right]
    =\haf(\Sigma).
\end{align}
Note that
\begin{align}
    \left|\frac{2^{n/2}}{(n/2)!}\left(h^\text{T}\Sigma h\right)^{n/2}\right|
    =\left|\frac{2^{n/2}}{(n/2)!}\left(\frac{n}{4}\hat{h}^\text{T}\Sigma \hat{h}\right)^{n/2}\right|
    =\left|\frac{n^{n/2}}{(n/2)!2^{n/2}}\left(\hat{h}^\text{T}\Sigma \hat{h}\right)^{n/2}\right|
    \leq \frac{(n\|\Sigma\|)^{n/2}}{(n/2)!2^{n/2}}.
\end{align}
Therefore, in general, the estimate by sampling $\bm{v}$ and using the estimator has an error bound given by Chernoff bound as
\begin{align}
    \text{Pr}\left[|\tilde{p}-\haf(\Sigma)|> \epsilon \frac{(n\|\Sigma\|)^{n/2}}{(n/2)!2^{n/2}}\right]\leq 2 \exp\left(-\frac{1}{2}\epsilon^2 T\right).
\end{align}
Here, $\|\Sigma\|$ is the spectral norm, i.e., the maximum singular value.

Our goal is to estimate the Fourier components of molecular vibronic spectra, which is given by
\begin{align}
    \tilde{G}(k)=\frac{\haf(\Sigma_{\bm{n}})}{Z\bm{n}!}.
\end{align}
Let us recall that the operator norm of the matrix $\Sigma_{\bm{\alpha},\bm{\beta}^*}$ is 1 (see Eq.~\eqref{eq:norm}).
Since a matrix's submatrix's operator norm is smaller than or equal to the matrix's norm, when the vector $\bm{n}$ is composed 0 or 1, $\Sigma_{\bm{n}}$'s operator norm is smaller than or equal to 1.
In addition, when $\bm{n}$ has a component larger than 1, we compute hafnian for a matrix obtained by repeating the row and column.
Nevertheless, when repeating the row and column the operator norm does not increase as much as $\bm{n}!$, i.e.,
\begin{align}
    \frac{\|\Sigma_{\bm{n}}\|}{\bm{n}!}\leq \|\Sigma\|.
\end{align}
Therefore, the estimate by sampling $\bm{x}$ and using the estimator has an error bound given by Chernoff bound as
\begin{align}
    \text{Pr}\left[\left|\tilde{q}-\frac{\haf(\Sigma)}{Z\bm{n}!}\right|\gtrsim \epsilon \frac{e^{n/2}}{\sqrt{\pi n}Z}\right]\leq 2 \exp\left(-\frac{1}{2}\epsilon^2 T\right),
\end{align}
where $Z=\sqrt{\prod_{i=1}^n\cosh{r_i}}$ and we have used Stirling's formula, $N!\approx \sqrt{2\pi N}(N/e)^N$ (more precisely, $\sqrt{2\pi N}(N/e)^N e^{\frac{1}{12N+1}}<N!<\sqrt{2\pi N}(N/e)^N e^{\frac{1}{12N}}$).
Thus, if $Z>e^{n/2}$, the estimation error is smaller than $\epsilon$ with exponentially small failure probability if we choose the number of samples as $T=O(1/\epsilon^2)$.

For nonzero mean, one may again employ a similar formula for $z\sim \mathcal{N}(\mu,\Sigma)$ \cite{kan2008moments}:
\begin{align}
    \lhaf(\Sigma_{\bm{n}})=\mathbb{E}_z\left[\prod_{i=1}^M z_i^{n_i}\right]=\sum_{v_1=0}^{n_1}\cdots \sum_{v_M=0}^{n_M}(-1)^{\sum_{i=1}^M v_i}\binom{n_1}{v_1}\cdots\binom{n_M}{v_M}\sum_{r=0}^{[n/2]}\frac{\left(\frac{h^\text{T}\Sigma h}{2}\right)^{r}(h\cdot \mu)^{n-2r}}{r!(n-2r)!},
\end{align}
where $n=n_1+\cdots+n_M$ and $h=(n_1/2-v_1,\dots,n_M/2-v_M)^\text{T}$ and $\Sigma_{\bm{n}}$ is obtained by replacing the diagonal elements of $\Sigma$ by $\mu$ and repeating $i$th row and column for $n_i$ times.
Again, let $n_i=1$ for all $i$'s. 
In this case, $M=n$, and 
\begin{align}
    \lhaf(\Sigma)&=\mathbb{E}_z\left[\prod_{i=1}^n z_i\right]=\sum_{v_1=0}^{1}\cdots \sum_{v_n=0}^{1}(-1)^{\sum_{i=1}^n v_i}\sum_{r=0}^{[n/2]}\frac{\left(\frac{h^\text{T}\Sigma h}{2}\right)^{r}(h\cdot \mu)^{n-2r}}{r!(n-2r)!} \\ 
    &=\frac{1}{2^{n}[n/2]}\mathbb{E}_{\bm{v},r}\left[(-1)^{\sum_{i=1}^n v_i}\frac{2^n[n/2]\left(\frac{h^\text{T}\Sigma h}{2}\right)^{r}(h\cdot \mu)^{n-2r}}{r!(n-2r)!}\right].
\end{align}
Here,
\begin{align}
    \left|\frac{2^n[n/2]\left(\frac{h^\text{T}\Sigma h}{2}\right)^{r}(h\cdot \mu)^{n-2r}}{r!(n-2r)!}\right|
    =\frac{[n/2]n^{n/2}\left(\hat{h}^\text{T}\Sigma \hat{h}\right)^{r}(\hat{h}\cdot \mu)^{n-2r}}{2^r r!(n-2r)!}
    \leq \frac{[n/2]n^{n/2}\|\Sigma\|^{r}\|\mu\|^{n-2r}}{2^r r!(n-2r)!}
\end{align}
Therefore, the Chernoff bound gives us
\begin{align}
    \text{Pr}\left[\left|p-\lhaf(\Sigma)\right|> \epsilon \frac{[n/2]n^{n/2}\|\Sigma\|^{r}\|\mu\|^{n-2r}}{2^r r!(n-2r)!}\right]\leq 2 \exp\left(-\frac{1}{2}\epsilon^2 T\right).
\end{align}
Especially when $\|\Sigma\|\leq 1$, which is the case that we are interested in,
\begin{align}
    \text{Pr}\left[\left|p-\lhaf(\Sigma)\right|> \epsilon \frac{[n/2]n^{n/2}\|\mu\|^{n-2r}}{2^r r!(n-2r)!}\right]\leq 2 \exp\left(-\frac{1}{2}\epsilon^2 T\right).
\end{align}
The error bound clearly depends on the mean vector $\mu$, which makes the bound more nontrivial than the hafnian case.

\section{Embedding a universal computing circuit into molecular vibronic spectra problem}\label{sec:BQP}
We first show that we can construct a BQP-complete spectra problem, which is essentially equivalent to simulating a universal computing circuit, in discrete-variable (DV) system.
To do that, we consider a quantum circuit $\hat{U}$ applied on $n$ qubits which consists of arbitrary single-qubit gates and CNOT gates with a polynomially large depth.
Now, we define the corresponding spectra problem as follows.
For a given circuit $\hat{U}$, $\Omega\in \{0,\dots,\Omega_\text{max}\}$, and $\bm{\omega}\in \mathbb{Z}_{\geq 0}^n$ with $\omega_i\leq O(\text{poly}(n))$, which are defined similarly as boson sampling cases, compute the spectra
\begin{align}
    G(\Omega)=\sum_{\bm{m}\in \{0,1\}^n}p(\bm{m})\delta(\Omega-\bm{\omega}\cdot \bm{m}).
\end{align}
We can easily show that one can approximate the single-qubit marginal probabilities if we can approximate the spectra.
We first choose $\bm{\omega}$ as $\omega_i=1$ for $i=1$ and $\omega_i=0$ otherwise.
Here, without loss of generality, we will compute the first qubit's marginal probabilities.
For this choice, each component of the spectra simply represents the marginal probability:
\begin{align}
    G(\Omega=1)=p(1),~~~
    G(\Omega=0)=p(0).
\end{align}
Hence, if we can approximate $G(\Omega)$, we can approximate the single-qubit marginal probabilities with the same accuracy.
Therefore, it proves that approximating the spectra within $O(\text{poly}(1/\epsilon))$ error in DV system is BQP-hard.
Also, since the spectra can be approximated with error in $O(\text{poly}(1/\epsilon))$ by simulating the quantum circuit, measuring the qubits and repeating sampling, the problem is in BQP, which shows that it is BQP-complete.

Using this property, we will construct a bosonic circuit whose spectra problem is also BQP-complete.
The key idea is to map a qubit-based circuit to a bosonic circuit using the Schwinger boson correspondence between qubit and boson states \cite{auerbach2012interacting}, which is similar to the one in Ref.~\cite{wei2010interacting}.
Specifically, one can map a single-qubit state $|x\rangle$ as
\begin{align}
    |x\rangle
    \iff (\hat{a}^\dagger)^x (\hat{b}^\dagger)^{1-x}|0\rangle,
\end{align}
where $|0\rangle$ is vacuum state.
Thus, one can map a single-qubit state to a single boson in a two-mode bosonic system, which corresponds to the dual-rail encoding in the Knill-Laflamme-Milburn protocol \cite{knill2001scheme}.
We call the two modes as a single site.
Similarly, an $n$-qubit state is mapped to a $2n$-mode state as
\begin{align}
    |\bm{x}\rangle \iff (\hat{a}_1^\dagger)^{x_1}(\hat{b}_1^\dagger)^{1-x_1}
    \cdots (\hat{a}_n^\dagger)^{x_n}(\hat{b}_n^\dagger)^{1-x_n}|0\rangle.
\end{align}
Furthermore, one can map an arbitrary single-qubit gate to an interaction between $\hat{a}$ and $\hat{b}$ by noting that any $2\times 2$ matrix can be decomposed by Pauli operators and using the following mapping
\begin{align}
    \hat{\sigma}_i^0 \iff \mathbb{1},~~~~
    \hat{\sigma}_i^1 \iff \hat{a}_i^\dagger\hat{b}_i+\hat{a}_i\hat{b}_i^\dagger,~~~~
    \hat{\sigma}_i^2 \iff i(\hat{a}_i\hat{b}_i^\dagger-\hat{a}_i^\dagger\hat{b}_i),~~~~
    \hat{\sigma}_i^3 \iff \hat{a}_i^\dagger\hat{a}_i-\hat{b}_i^\dagger\hat{b}_i.
\end{align}
Here, it is worth emphasizing that all the operations mapped from the Pauli operators preserve the number of photons and restrict the number of photons to a single photon for each site.
Notice that the operators in bosonic system are not unitary operators in the full Hilbert space.
However, since the subspace that we need to consider is confined by the states spanned by the single-photon states per each site,
within this confined subspace, one can easily verify that the mapped operators in bosonic systems are guaranteed to be unitary.
As a remark, for single-qubit gates, instead of directly mapping the unitary operation, one can map the Hamiltonian of the operation using the Euler decomposition, whose unitary operator will be decomposed into beam splitters and phase shifters.

For two-qubit gates as well, we can decompose arbitrary operations using Pauli operators as
\begin{align}
    \hat{U}=\sum_{\mu,\nu=0}^3 u_{\mu,\nu} \hat{\sigma}_i^{\mu}\otimes \hat{\sigma}_j^{\nu},
\end{align}
and map this to bosonic operators as single-qubit cases.
Again, all the operations mapped from the Pauli operators preserve the number of photons and restrict the number of photons to a single photon for each site.
Therefore, we can map arbitrary input states, single-qubit gates, and two-qubit gates in an $n$-qubit system to a $2n$-mode bosonic system.
Hence, it shows that we can construct a BQP-complete problem in bosonic system as well.

Now, we emphasize that the constructed bosonic circuit is different from Fock-state boson sampling for various aspects.
First, although the circuit preserves the number of photons, we do not expect the circuit can be simulated by linear optics unless Fock-state boson sampling constitutes universal quantum computation.
Therefore, the implementation of the circuit requires non-linear effects that are beyond quadratic hamiltonians, i.e., beyond Gaussian operations.
Second, the constructed circuit guarantees that the number of photons in each site composed of $\hat{a}_i$ and $\hat{b}_i$ is always occupied by a single photon, which is clearly not the case for Fock-state boson sampling.
It implies not only that the constructed circuit is distinct from Fock-state boson sampling but also that this is not the most general bosonic circuit that we can consider.
Therefore, general molecular vibronic spectra problems that allow non-linear operations and multiphoton cases, which include the constructed circuit as a particular example, are BQP-hard.

\bibliography{reference.bib}